\documentclass[fleqn,usenatbib]{mnras}

\usepackage{newtxtext,newtxmath}

\usepackage[T1]{fontenc}
\usepackage{ae,aecompl}

\usepackage{graphicx}	
\usepackage{amsmath}	
\usepackage{amssymb}	
\usepackage{titlesec}
\usepackage{epstopdf}
\usepackage{booktabs}
\usepackage{pdflscape}	
\usepackage[authoryear]{natbib}
\usepackage[labelformat=parens,labelsep=none,subrefformat=parens]{subcaption}
\usepackage{amsmath}
\usepackage{multirow}
\usepackage{color}

\usepackage{array}
\newcommand{\PreserveBackslash}[1]{\let\temp=\\#1\let\\=\temp}

\newcommand{\hii}{H\,{\textsc{ii}}}
\newcommand{\hi}{H\,{\textsc{i}}}
\newcommand{\radms}{rad\,m$^{-2}$}
\newcommand{\halpha}{H\,$\alpha$}
\graphicspath{{./}{figures/}}

\title[Faraday tomography towards Sh2-27]{Through thick or thin: Multiple components of the magneto-ionic medium towards the nearby H\,{\Large \textbf{II}} region Sharpless 2-27 revealed by Faraday tomography}

\author[Alec J. M. Thomson et al.]{
\newauthor Alec J. M. Thomson,$^{1}$\thanks{E-mail: \url{alec.thomson@anu.edu.au}}
T.L. Landecker,$^{2}$
John M. Dickey,$^{3}$
N.M. McClure-Griffiths,$^{1}$
\newauthor M. Wolleben,$^{4}$
E. Carretti,$^{5}$
A. Fletcher,$^{6}$
Christoph Federrath,$^{1}$
A. S. Hill,$^{2,7,8}$
\newauthor S. A. Mao,$^{9}$
B. M. Gaensler,$^{10,11}$
M. Haverkorn,$^{12}$
S. E. Clark,$^{13}$\thanks{Hubble Fellow}
C. L. Van Eck,$^{10}$
\newauthor and J. L. West$^{10}$
\\
$^{1}$Research School of Astronomy and Astrophysics, Australian National University, Canberra, ACT 2611, Australia\\
$^{2}$National Research Council Canada, Dominion Radio Astrophysical Observatory, P.O. Box 248, Penticton, British Columbia, V2A 6J9, Canada\\
$^{3}$School of Natural Sciences, Private Bag 37, University of Tasmania, Hobart, TAS, 7001, Australia\\
$^{4}$Skaha Remote Sensing Ltd., 3165 Juniper Drive, Naramata, British Columbia V0H 1N0, Canada\\
$^{5}$INAF - Istituto di Radioastronomia, Via P. Gobetti 101, I-40129, Bologna, Italy\\
$^{6}$School of Mathematics, Statistics and Physics, Newcastle University, Newcastle-upon-Tyne, NE13 7RU, UK\\
$^{7}$Department of Physics and Astronomy, University of British Columbia, Vancouver, BC V6T 1Z1, Canada\\
$^{8}$Space Science Institute, Boulder, CO 80301, USA\\
$^{9}$Max-Planck-Institut fur Radioastronomie, 53121 Bonn, Germany\\
$^{10}$Dunlap Institute for Astronomy and Astrophysics, University of Toronto, 50 St. George Street, Toronto, ON M5S 3H4, Canada\\
$^{11}$Department of Astronomy and Astrophysics, University of Toronto, Toronto, ON M5S 3H4, Canada\\
$^{12}$Department of Astrophysics/IMAPP, Radboud University, P.O. Box 9010, NL-6500 GL Nijmegen, The Netherlands\\
$^{13}$Institute for Advanced Study, 1 Einstein Drive, Princeton, NJ 08540, USA
}
\date{Accepted XXX. Received YYY; in original form ZZZ}

\pubyear{2019}

\begin{document}
\label{firstpage}
\pagerange{\pageref{firstpage}--\pageref{lastpage}}
\maketitle

\begin{abstract}
Sharpless 2-27 (Sh2-27) is a nearby \hii\ region excited by $\zeta$Oph. We present observations of polarized radio emission from 300 to 480\,MHz towards Sh2-27, made with the Parkes 64\,m Radio Telescope as part of the Global Magneto-Ionic Medium Survey. These observations have an angular resolution of $1.35^{\circ}$, and the data are uniquely sensitive to magneto-ionic structure on large angular scales. We demonstrate that background polarized emission towards Sh2-27 is totally depolarized in our observations, allowing us to investigate the foreground. We analyse the results of Faraday tomography, mapping the magnetised interstellar medium along the 165\,pc path to Sh2-27. The Faraday dispersion function in this direction has peaks at three Faraday depths. We consider both Faraday thick and thin models for this observation, finding that the thin model is preferred. We further model this as Faraday rotation of diffuse synchrotron emission in the Local Bubble and in two foreground neutral clouds. The Local Bubble extends for 80\,pc in this direction, and we find a Faraday depth of $-0.8 \pm 0.4\,$ \radms. This indicates a field directed away from the Sun with a strength of $-2.5\pm1.2\,\mu$G. The near and far neutral clouds are each about 30\,pc thick, and we find Faraday depths of $-6.6\pm0.6\,$\radms\ and  $+13.7\pm0.8\,$\radms, respectively. We estimate that the line-of-sight magnetic strengths in the near and far cloud are $B_{\parallel, \text{near}} \approx -15\,\mu\text{G}$ and $B_{\parallel, \text{far}} \approx +30\,\mu\text{G}$. Our results demonstrate that Faraday tomography can be used to investigate the magneto-ionic properties of foreground features in front of nearby \hii\ regions.
\end{abstract}

\begin{keywords}
polarization -- ISM: magnetic fields -- (ISM:) \hii\ regions
\end{keywords}



\section{Introduction} \label{sec:intro}
Magnetic fields are crucial dynamical drivers in the Galactic interstellar medium (ISM). They are responsible for injecting significant energy into the ISM \citep{Heiles2012,Beck2013,Beck2016}. Magnetic fields play roles in star formation and turbulent gas flows \citep{Padoan2011,Federrath2012,Federrath2015}, and also have profound consequences for the initial mass function of stars \citep{Offner2014, Federrath2014}. Despite their importance, much remains unknown regarding both the magnitude and structure of these magnetic fields. This has arisen from the general difficulty in measuring the strength of structure of magnetic fields in the ISM.

Radio spectro-polarimetry is one of the most effective ways to study interstellar magnetic fields~\citep{Han2017}. Linearly polarized emission is produced within the Milky Way by relativistic electrons emitting synchrotron radiation as they orbit around magnetic fields. 
At radio frequencies this emission suffers Faraday rotation as it propagates towards the observer through the magneto-ionic medium (MIM). Thus, observations of Galactic polarized radio emission contain a wealth of information on the Milky Way's magneto-ionic structure.

Faraday rotation causes the polarization angle ($\chi$) of an electromagnetic wave to rotate from an initial angle ($\chi_0$) at wavelength $\lambda$:
\begin{equation}
\chi(\lambda^2) = \chi_0 + \lambda^2\phi,
\label{eqn:pa}
\end{equation}
where $\phi$ is the Faraday depth \citep{Burn1966,Brentjens2005}:
\begin{equation}
\phi(d) \equiv 0.812\int_{d}^{0}n_e(r)B_\parallel(r) dr \left[ \text{\radms}\right],
\label{eqn:faradaydepth}
\end{equation}
and $n_e$ is the thermal electron density in cm$^{-3}$, $B_\parallel$ is the line-of-sight (LOS) component of the magnetic field in $\mu$G, and $dr$ is the incremental distance along the LOS in pc to a source at distance $d$. In the case of a single rotating region in front of a polarized source, referred to as a `Faraday screen', the Faraday depth is equivalent to the rotation measure (RM):
\begin{equation}
\text{RM} \equiv \frac{d\chi}{d(\lambda^2)} \left[\text{\radms}\right].
\label{eqn:rm}
\end{equation}
We follow the definitions of \citet{Brentjens2005} throughout, we quantify Faraday rotation using Faraday depth, and we refer to RMs from extragalactic sources. Due to the strong wavelength dependence, low-frequency radio observations of polarized emission are very sensitive for measuring Faraday rotation in the magneto-ionic medium (MIM). The determination of the Faraday depth from Galactic synchrotron emission is non-trivial, however, due both to the complexity of the Galactic MIM and the mixing of emission and Faraday rotation in the same volume. This can be overcome by mapping polarization across many frequency channels in a technique called `Faraday tomography'. We outline this technique in Section~\ref{sec:methods}.

The large angular scales of diffuse Galactic polarized emission calls for global radio spectro-polarimetric survey. The Global Magneto-Ionic Medium Survey \citep[GMIMS,][]{Wolleben2008} was devised specifically to probe the MIM of the Milky Way. This survey will ultimately measure diffuse polarized emission across the entire sky from 300\,MHz to 1.8\,GHz using single-dish telescopes, giving excellent sensitivity to a wide range of Faraday structures. Results from the GMIMS high-band North \citep[GMIMS-HBN,][]{Wolleben2010a}, taken with the DRAO 26\,m telescope, have been used directly to investigate the magneto-ionic properties of a nearby \hi\ shell \citep{Wolleben2010}, the North Polar Spur \citep{Sun2015}, and the Fan Region \citep{Hill2017}, and they are incorporated into other work analysing all-sky emission \citep[e.g.][]{Dickey2018,Zheng2017}.

The nearby \hii\ region Sharpless 2-27 (Sh2-27) appears in various radio polarization observations. Sh2-27 surrounds the star $\zeta$Oph which is located at $[l,b]\sim[6.3^{\circ},+23.6^{\circ}]$~\citep{VanLeeuwen2007}. The region subtends about $10^{\circ}$ on the sky and is readily identifiable in \halpha\ images. \hii\ regions are highly ionized regions of the ISM, and thus have a greater thermal electron density over the typical Galactic warm neutral medium \citet{Ferriere2001}. In the presence of magnetic fields \hii\ regions have a strong effect on observations of radio polarization~\citep[e.g.][]{Gaensler2001}. At 2.3\,GHz in the S-band Polarization All Sky Survey \citep[S-PASS,][]{Carretti2019} Sh2-27 has been identified as a Faraday screen, modulating the polarization angle but not producing polarized emission itself~\citep{Robitaille2017, Robitaille2018, Iacobelli2014}. In polarization observations at 1.4\,GHz, such as GMIMS-HBN, Sh2-27 can be identified as a depolarizing region. \citet{Wolleben2010} used the depolarization of Sh2-27 to constrain the distance of polarized emission through a nearby \hi\ shell. The magneto-ionic properties of Sh2-27 were directly investigated by \citet{Harvey-Smith2011a} using the NVSS catalogue of point-source RMs~\citep{Taylor2009}. This region stands out in the \citet{Taylor2009} catalogue, and derivative maps such as \citet{Oppermann2012,Oppermann2015}, due to high values of RM from extragalactic sources seen through it.

In this paper we present results from the low-band Southern Global Magneto-Ionic Medium Survey (GMIMS-LBS) towards Sh2-27. Using these data we are able to isolate a column of foreground MIM for analysis with Faraday tomography. The distance to Sh2-27 is known to be $\sim180\,$pc~\citep{GaiaCollaboration2016, GaiaCollaboration2018}, which means we are able to map results from polarization observations within that distance. We provide additional background and definitions we use that are specific to radio polarimetry in Section~\ref{sec:methods}.
We describe the GMIMS-LBS observations in Section~\ref{sec:obs}, including the application of Faraday tomography. In Section~\ref{sec:results} we present the results of these observations towards Sh2-27 and show that it is depolarizing the background emission in the GMIMS-LBS band. We conclude that Sh2-27 is acting as a `depolarization wall' for extended structures, and can therefore be used to constrain distances in Faraday tomography. We describe the structure in the GMIMS-LBS Faraday depth cubes towards Sh2-27 in Section~\ref{sec:cubes}. We analyse how this structure maps to distance along the LOS in Section~\ref{sec:analysis}. In Section~\ref{sec:modelism} we consider a Faraday thin interpretation in combination with data on the local ISM to both reconstruct the magnetic field structure and estimate the magnetic strength along the LOS. In Section~\ref{sec:modelcube} we consider an alternate model using Faraday thick structures. We discuss our results in Section~\ref{sec:discuss}, and provide a summary and conclusion in Section~\ref{sec:summary}.

\section{Background}\label{sec:methods}
\subsection{Faraday Tomography}
It is highly unlikely that any given LOS in the Galaxy would be as simple as a Faraday screen. With this in mind, the technique of Faraday tomography (also known as RM synthesis) \citep{Burn1966, Brentjens2005, Heald2009} was developed. This method applies a discrete Fourier transform to the complex polarization as a function of $\lambda^2$. The primary result of this technique is the Faraday dispersion function ($F(\phi)$), the polarized flux as a function of Faraday depth. This function is spectral in nature, and we refer to it as the Faraday spectrum. The output parameters of Faraday tomography are set by the behaviour of the `RM spread function' (RMSF). The effective resolution of the Faraday spectra ($\delta\phi$) is given by the width of the RMSF at full-width of half maximum (FWHM) \citep{Brentjens2005}:
\begin{equation}
	\delta\phi \approx \frac{2\sqrt{3}}{\Delta\lambda^2}
	\label{eqn:faradayres}
\end{equation}
where $\Delta\lambda^2=\lambda_{\text{max}}^2 - \lambda_{\text{min}}^2$ is the bandwidth in $\lambda^2$-space, and $\lambda_{\text{max}}^2$ and $\lambda_{\text{min}}^2$ are the maximum and minimum observed $\lambda^2$, respectively. The largest observable value of Faraday depth ($\phi_{\text{max}}$) is set by the width of the observed $\lambda^2$ channels ($\delta\lambda^2$):
\begin{equation}
	\phi_{\text{max}} \approx \frac{\sqrt{3}}{\delta\lambda^2}
\end{equation}
Finally, the smallest observed $\lambda^2$ sets the maximum scale observable in Faraday depth space:
\begin{equation}
	\phi_{\text{max-scale}} \approx \frac{\pi}{\lambda_{\text{min}}^2}
\end{equation}
Sources that produce a broad feature in the Faraday spectrum are referred to as `Faraday thick'. Specifically, a source is `thick' if $\lambda^2\Delta\phi\gg1$, where $\Delta\phi$ is the extent of the source in $F(\phi)$ observed at $\lambda^2$ \citep{Brentjens2005}. Such features can be modelled as a mixture of a coherent and turbulent magnetic field that produces both synchrotron emission and Faraday rotation of background polarized emission~\citep{Burn1966, Sokoloff1998}. Conversely, a feature is Faraday thin if $\lambda^2\Delta\phi\ll1$. Faraday thin features can be modelled as a $\delta$ function in the Faraday spectrum.

Observational restrictions on wavelength coverage have a strong effect on Faraday tomography. These effects can be mitigated using deconvolution techniques. Currently, the most popular algorithm is \verb|RM-CLEAN| \citep{Heald2009}, which replaces the `dirty' RMSF with a smooth Gaussian restoring beam. This reduces the effect of sidelobes that are present in the `dirty' Faraday spectra.

\subsection{Depolarization}
Depolarization is a common feature of almost all radio polarization observations, with the exception of polarized emissions from pulsars. This effect can occur through three primary mechanisms~\citep{Burn1966, Tribble1991, Sokoloff1998}: depth, beam, and bandwidth depolarization. Depth depolarization refers to the effect of Faraday thick sources in $\lambda^2$ space. Such sources lose polarized flux as a function of $\lambda^2$. Beam and bandwidth depolarization arise from observational parameters. In the former case, the variation of Faraday depth occurs spatially within the beam of the telescope. Bandwidth depolarization occurs when significant Faraday rotation occurs within one frequency channel.

In low-frequency observations depolarization features become far more common and are often associated with ionised regions of the ISM, such as \hii\ regions. As these features depolarize emission from behind them, they can be used as distance indicators in radio polarization observations. 

Despite their higher Faraday resolution, low-frequency observations can face an issue by not observing polarized flux at short $\lambda^2$. The result of missing this emission is that sources with a Faraday thickness greater than $\phi_{\text{max-scale}}$ are `resolved out', whereby broad features are lost leaving only narrow features present in Faraday depth space. In practice this can give rise to an ambiguity between a Faraday thick feature or a number of Faraday thin features.

A `depolarization wall' \citep{Hill2018} is a form of spatially discrete depolarization. Whilst conceptually similar to the `polarization horizon' \citep{Uyaniker2003}, a depolarization wall arises when a specific and discrete depolarising object (such as an \hii\ region) lies along the LOS. When a LOS passes through a wall the background polarized emission is totally depolarized. Whether or not an object acts as a wall in a given observation will depend on both the observed $\lambda^2$ and the angular resolution. Polarization walls have a great utility for analysing results of Faraday tomography. Despite the large amount of information contained within Faraday spectra, mapping that structure to physical space is challenging. If the distance to a depolarization wall can be determined, however, that places a constraint on the distance along which the observed Faraday structure occurs. This is highly analogous to the use of \hii\ regions as free-free absorbers of Galactic synchrotron emission \citep[e.g.][]{Nord2006,Su2018}.

\section{Observations} \label{sec:obs}
\subsection{GMIMS Low-Band South}
Recently we completed GMIMS-LBS with the Parkes 64\,m telescope. A complete description of these observations is provided in Wolleben et al. (submitted). These observations measure diffuse polarized emission (Stokes $I$, $Q$, and $U$) across the entire Southern sky from 300\,MHz to 480\,MHz with a spectral resolution of 0.5\,MHz.

Here we analyse the Faraday spectral cubes from this survey. These spectra have been deconvolved using \verb|RM-CLEAN|~\citep{Heald2009}. We summarise the properties of these data, including the parameters resulting from Faraday tomography, in Table~\ref{tab:survey}. The long wavelengths and high spectral resolution result in a unique property for this survey: a very fine Faraday resolution of $\delta\phi = 6.2\,$\radms, smaller than the Faraday max-scale of the survey.  This is the first large-scale sky survey with ${\phi_\text{max-scale}}>{{\delta}{\phi}}$ at frequencies above 250\,MHz. This property means that \textit{only} features that are broader than $\phi_{\text{max-scale}}$ will be resolved out. Without this property, the observed spectra become more complex \citep{Dickey2018} and their interpretation more difficult.

It is also important to consider the behaviour of noise in Faraday spectra. The RMS noise in the Stokes $Q$ and $U$ spectra is $\sigma_{QU} = 60\,$mK. We primarily consider the absolute value of the Faraday dispersion function, which represents the polarized intensity. When analysing the polarized intensity the variance ($\sigma_{\text{PI}}$) is given by a Rayleigh distribution~\citep{Wardle1974, Heald2009}:
\begin{equation}
	\sigma_{\text{PI}} = \sqrt{\frac{4-\pi}{2}}\sigma_{QU} \approx 0.66\sigma_{QU},
\end{equation}
in the low signal-to-noise limit. For increasing signal-to-noise the variance approaches a Gaussian distribution and $\sigma_{\text{PI}} = \sigma_{QU}$. 

\begin{table}
	\centering
	\caption{Summary of the observational parameters of the GMIMS-LBS \citep[Wolleben et al. submitted,][]{Dickey2018}. $^a$ -- This range is determined by the high and low signal-to-noise limits. $^b$ -- We select these values during Faraday tomography.} \label{tab:survey}

	\begin{tabular}{l l c c}
		\toprule
		\midrule
		Survey parameter 					& Symbol  & min. 					& max. 		\\
		\midrule
		Declination [$^{\circ}$] 				& $\delta$ & $-90$ 				& $+20$ 	\\
		Beamwidth [$\arcmin$] 		& & 79.4 					& 83.6 		\\
		Frequency [MHz] 					& $f$ & 300.25 				& 479.75 	\\
		Frequency resolution [MHz] 			& $\delta f$ & \multicolumn{2}{c}{0.5} \\
		Wavelength-squared [m$^{2}$] 		& $\lambda^2$ & 0.391 				& 0.999 	\\
		$\lambda^2$ bandwidth [m$^{2}$] 	& $\Delta\lambda^2$ & \multicolumn{2}{c}{0.608} \\
		$\lambda^2$ resolution [m$^{2}$] 	& $\delta\lambda^2$ & \multicolumn{2}{c}{$3.32\times10^{-3}$} \\
		Stokes $Q$ and $U$ RMS noise [mK] 	& $\sigma_{QU}$ 	& \multicolumn{2}{c}{60} \\
		PI RMS noise$^a$ [mk]					& $\sigma_{\text{PI}}$ 	& 39 & 60 \\
		Faraday resolution [\radms] 		& $\delta\phi$ & \multicolumn{2}{c}{6.2}	\\
		Max. Faraday depth [\radms] 		& $\phi_{\text{max}}$ & \multicolumn{2}{c}{$1.3\times10^3$} \\
		Faraday max. scale [\radms] 		& $\phi_{\text{max-scale}}$ & \multicolumn{2}{c}{8.0} \\
		$\phi$ range$^b$ [\radms] 				& & $-100$ 				& $+100$ 	\\
		$\phi$ sampling$^b$ [\radms] 			& & \multicolumn{2}{c}{0.5} \\
		\bottomrule
	\end{tabular}
\end{table}

\subsection{Complementary data}
We use a number of other datasets to complement our GMIMS-LBS observation. \citet{Finkbeiner2003} combines data from the Virginia Tech Spectral Line Survey~\citep[VTSS,][]{Dennison1998}, the Southern H-Alpha Sky Survey~\citep[SHASSA,][]{Gaustad2001} and the Wisconsin H-alpha mapper~\citep[WHAM,][]{Haffner2003} to produce an all-sky \halpha\ intensity image with a resolution of 6\arcmin. We use these data to identify Sh2-27 and other \hii\ regions around it.

The \citet{Taylor2009} catalogue provides measurements of RM towards extragalactic point sources as measured by the Very Large Array (VLA). These data are derived from NRAO VLA Sky Survey \citep[NVSS,][]{Condon1998}, and provide a source density of $\sim1\,$deg$^{-2}$. Since these data were taken at L-band, and with 45\arcsec resolution, they are far less susceptible to depolarization effects. We are therefore able to investigate the Faraday rotation through Sh2-27 with these data.

The STructuring by Inversion the Local Interstellar Medium project\footnote{\href{https://stilism.obspm.fr/}{https://stilism.obspm.fr/}, version 4.1, accessed October 2018} \citep[STILISM,][]{Lallement2014,Capitanio2017,Lallement2018} provides information on the three-dimensional structure of the nearby ISM. These data are produced using dust reddening of starlight \citep[e.g.][]{Vergely2010a,Lallement2014,Green2014,Capitanio2017,Green2018,Lallement2018}, with stellar parallax distances from \textit{Gaia}, to map dust features in the nearby ISM. We use the data cube from this project, which covers a 4\,kpc by 4\,kpc by 600\,pc grid around the Sun.

\section{Results} \label{sec:results}
\subsection{Depolarization from Sh2-27}

\begin{figure}
	\centering
	\includegraphics[width=\columnwidth]{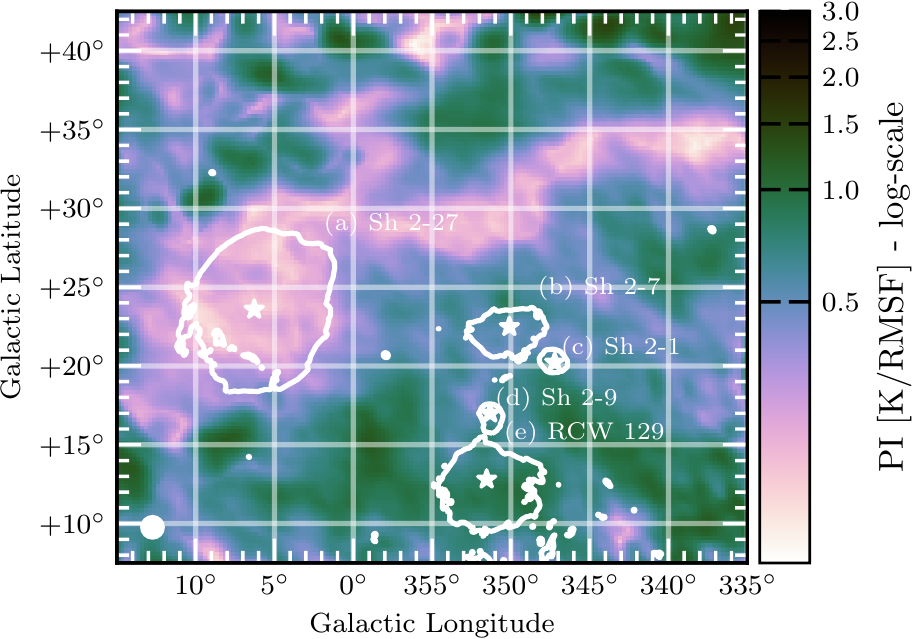}
	\caption{The peak PI in the Faraday cube towards Sh2-27. Contours are \halpha\ intensity from \citet{Finkbeiner2003} at 30\,R. We label the five visible \hii\ regions, and their corresponding central stars (white stars), in this region as: (a) -- Sh2-27 / $\zeta$Oph, (b) -- Sh2-7 / $\delta$Sco, (c) -- Sh2-1 / $\pi$Sco, (d) -- Sh2-9 / $\sigma$Sco, (e) -- RCW 129 / $\tau$Sco. We show the beam as a white circle in the lower-left corner. We note that in \halpha\ there are four other nearby \hii\ regions that appear close on the sky to Sh2-27. In contrast to Sh2-27, these \hii\ regions have no discernible effect on the polarization data. We identify a depolarization wall that occurs approximately within the \halpha\ contour of Sh2-27. We further find that the depolarized feature extending horizontally across this map is a depolarization canal.}
	\label{fig:peakpi}
\end{figure}

Polarized intensity is very low in GMIMS-LBS towards Sh2-27. The depolarizing effect of Sh2-27 in our data can be seen in Figure~\ref{fig:peakpi}, which shows the peak polarized intensity from the \verb|CLEAN| Faraday spectra in the region towards Sh2-27. We also show the combined SHASSA and WHAM \halpha\ intensity from \citet{Finkbeiner2003} as white contours. We identify two important features from this map. First, while the area towards Sh2-27 is clearly reduced in polarized intensity with respect to the surrounding emission, the polarized intensity is well above the noise (60\,mK). Second, a strong but narrow depolarization feature extends out to the right from the edge of the Sh2-27's depolarization region. We will address these features in turn with respect to several depolarization mechanisms.

GMIMS-LBS is able to probe magneto-ionic effects in great detail due to the long wavelengths observed. Consequently, these observations are also more sensitive to depolarization features. A Faraday depth of about $\pm940\,$\radms\ would be required to completely depolarize our lowest frequency observation through bandwidth depolarization. Such extreme values are rarely observed away from the Galactic plane. We therefore do not expect bandwidth depolarization to affect our observations. 

Given the large beam of GMIMS-LBS (81\,arcmin at 300\,MHz), beam depolarization is likely to be a significant effect. We quantify the beam depolarization towards Sh2-27 using point-source RMs. These values probe Faraday rotation along the entire LOS out to the edge of the Galaxy, thus allowing the investigation of the intervening ISM. 

\begin{figure*}
	\centering
	\begin{subfigure}[b]{0.49\textwidth}
		\includegraphics[width=\textwidth]{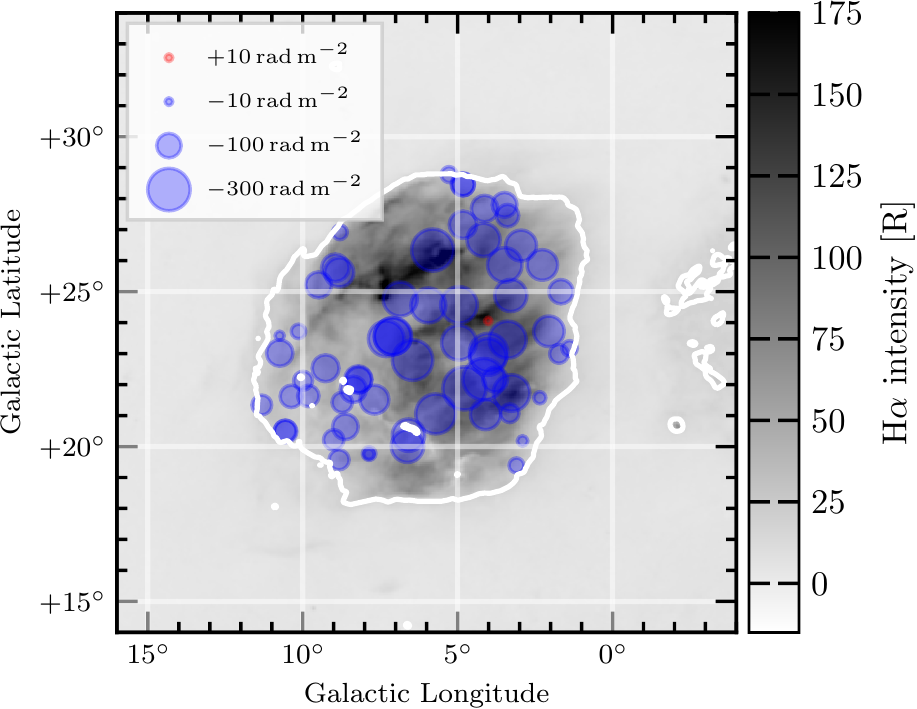}
		\caption{}
		\label{fig:nvsscor}%
	\end{subfigure}
	~ 
	\begin{subfigure}[b]{0.4\textwidth}
		\includegraphics[width=\textwidth]{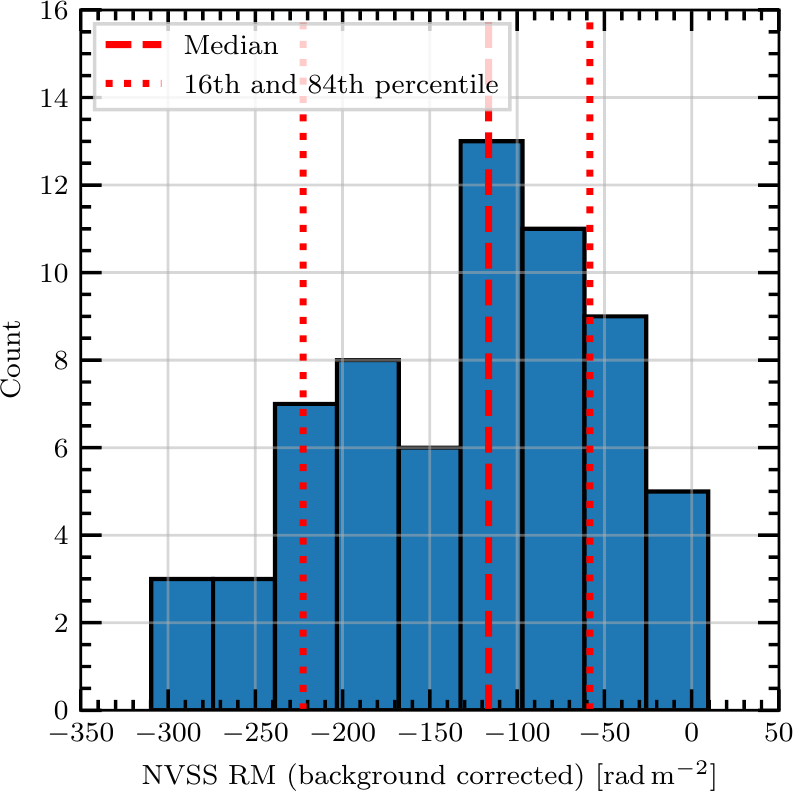}
		\caption{}
		\label{fig:nvssdist}%
	\end{subfigure}
	\caption{The \citet{Taylor2009} RMs towards Sh2-27. Here, we apply the selection criteria and background correction of \citet{Harvey-Smith2011a}. \subref{fig:nvsscor} The spatial distribution of RMs on Sh2-27. \subref{fig:nvssdist} The histogram of the RM distribution towards Sh2-27. We also show the median RM (dashed line), and 16th and 84th percentiles (dotted lines). We use these data to demonstrate that Sh2-27 is a depolarization wall to the diffuse emission measured by GMIMS-LBS. The high RM values shown here are not detected in our Faraday spectra as polarized emission from behind the \hii\ region is totally depolarized.}
\end{figure*}

Here we apply a similar analysis to \citet{Harvey-Smith2011a}, but instead we will obtain the variation in Faraday depth across Sh2-27, and thus estimate the beam depolarization in GMIMS-LBS using the \citet{Taylor2009} catalogue. We adopt the same boundary conditions and background RM correction as \citet{Harvey-Smith2011a}, given in their Table 2. This results in 65 background-corrected RMs through Sh2-27, which we show in Figure~\ref{fig:nvsscor}. We also show the distribution of these RMs in Figure~\ref{fig:nvssdist}. From these RMs we find a median value of $-166\,$\radms\ and a standard deviation of $\sigma_{\text{RM}}=78$\,\radms. To analyse how $\sigma_{\text{RM}}$ changes across angular scales we compute the second-order structure function ($\text{SF}_\text{RM}$) of the RMs on Sh2-27, as defined by \citet{Haverkorn2004}:
\begin{equation}
	\text{SF}_\text{RM} (\Delta\theta) = \langle [\text{RM}(\theta) - \text{RM}(\theta + \Delta\theta)]^2 \rangle,
\end{equation}
where $\Delta\theta$ is the angular distance on the sky between two LOS, and $\left\langle  \dots\right\rangle$ represents the average on all pairs  of separation $\Delta \theta$. We estimate the errors in the structure function by utilising Monte-Carlo error propagation. Assuming that the errors in the \citet{Taylor2009} RMs are Gaussian distributed, we take 1000 samples of a Gaussian distribution for each RM on Sh2-27 and propagate the entire distribution through the $\text{SF}_\text{RM}$ computation. We find that the function remains flat from the angular scale of Sh2-27 ($\sim10^{\circ}$) to scales smaller than the beamwidth of our observations. We can therefore expect that the variation in RM as computed across the entire Sh2-27 region will be about the same as the variation within the GMIMS-LBS beam.

We estimate that the variance in Faraday depth due to Sh2-27 can be related to the variation in RM by:
\begin{equation}
\sigma^2_{\text{RM}} = \sigma^2_\text{\hii} + \sigma^2_{\text{gal}} + \sigma^2_{\text{exgal}}  + \sigma^2_\text{err}
\end{equation}
where $\sigma_\text{\hii}$ is the variation in Faraday depth caused by turbulent structures in the \hii\ region, $\sigma_{\text{gal}}\approx 8 / \sin{(b)} \approx 20\,$\radms\ \citep{Schnitzeler2010} is the variation along the rest of the LOS through the Galaxy, $\sigma_{\text{exgal}}\approx6\,$\radms\ \citep{Schnitzeler2010} is the variation in RM due to contribution from the intrinsic Faraday rotation of the extragalactic source, and $\sigma_\text{err}=10.1\pm0.4$ is the measurement error in RM. In this way we estimate the variation in Faraday depth of Sh2-27 to be $\sigma_\text{\hii}\approx74\pm1\,$\radms. The degree of beam depolarization can be quantified by either the \citet{Burn1966} depolarization law, or by the \citet{Tribble1991} depolarization law if the depolarization (compared to the intrinsic polarisation fraction) is $<0.5$: 
\begin{align}
	\text{DP}_\text{Burn} &= e^{-2\sigma^2\lambda^4}\\
	\text{DP}_\text{Tribble} &= \frac{1}{2\sqrt{2}\sqrt{N}\sigma\lambda^2}
\end{align}
where $\text{DP}$ is the depolarization fraction (the ratio of observed to intrinsic PI), $\sigma$ is the variation in Faraday depth, $\lambda$ is the observed wavelength, and $N$ is the number of independent, randomly varying areas within the beam. Across our band, the Burn depolarization factor is $<\exp{(-1700)}$ and the Tribble depolarization factor is $<1/(130\sqrt{N})$ ($<0.008$ for $N=1$). In either case, the emission behind Sh2-27 is strongly beam depolarized in our survey. We find, however, a significant polarized signal towards Sh2-27. Since an \hii\ region does not produce polarized emission itself we are able to proceed treating Sh2-27 as a `depolarization wall' and we conclude that the polarized emission that we observe must arise between the Sun and Sh2-27.

\citet{Hill2018} does note, however, that it is possible for polarization to make its way through a depolarizing volume, such as an \hii\ region, using a semi-analytic mock observation matched to GMIMS-LBS. Their model included a lower-density \hii\ region than Sh2-27. We ran a version of their model with a density and magnetic field which matches estimates for Sh2-27 \citep{Harvey-Smith2011a}. Some polarized radiation does leak through at the Faraday depth of the \hii\ region in the model, but the polarized intensity is $\lesssim 10\%$ of the background polarized intensity. In the model, there are components of the Faraday spectrum at Faraday depths comparable to what would be observed for background sources; we do not see components at the Faraday depths seen by \citet{Harvey-Smith2011a}, so the depolarization may be more wall-like than in the \citet{Hill2018} model.

\begin{figure*}
	\centering
	\begin{subfigure}[b]{0.49\textwidth}
		\includegraphics[width=\textwidth]{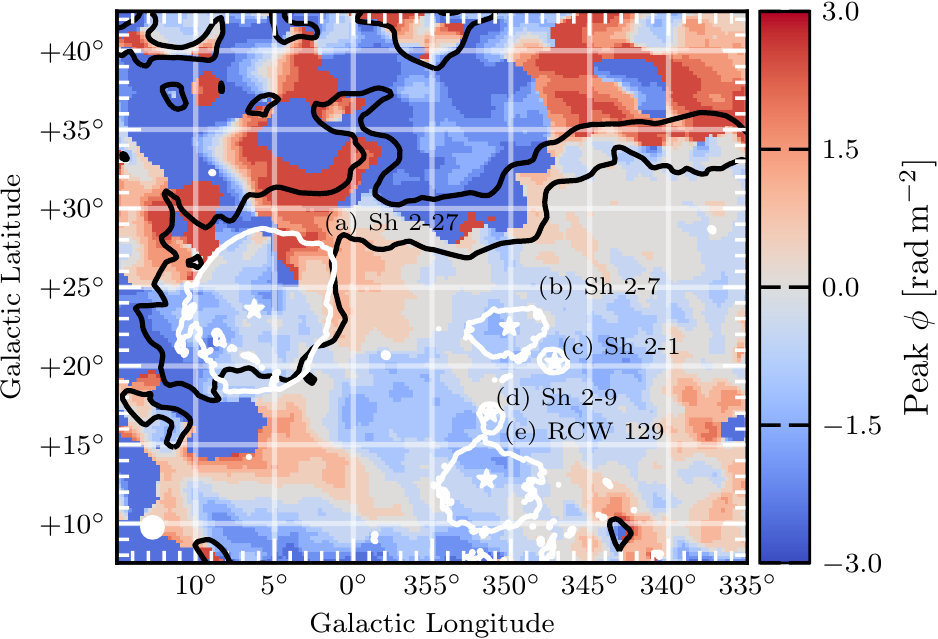}
		\caption{}
		\label{fig:peakrm}%
	\end{subfigure}
	~ 
	\begin{subfigure}[b]{0.49\textwidth}
		\includegraphics[width=\textwidth]{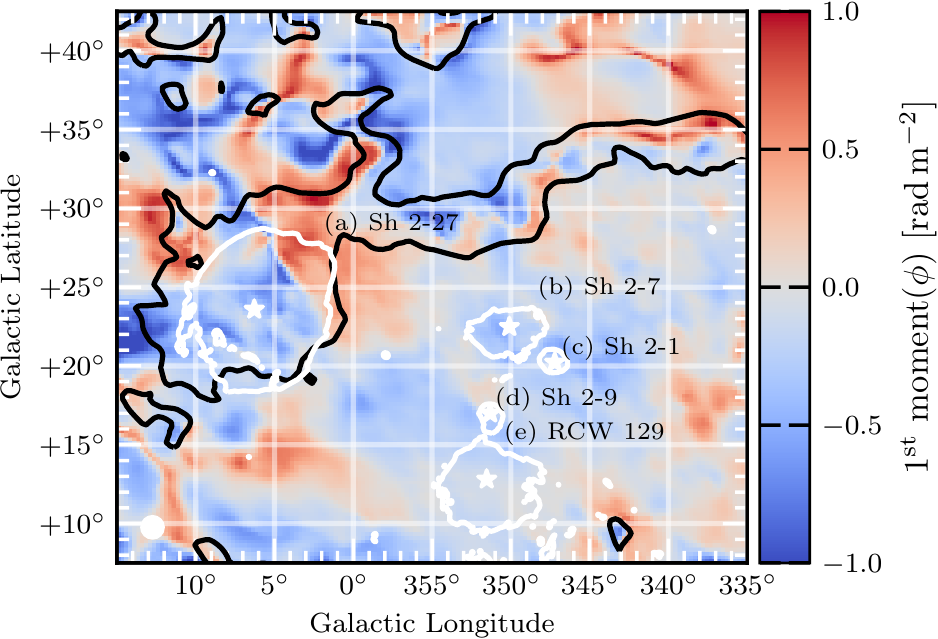}
		\caption{}
		\label{fig:peakmom}%
	\end{subfigure}
	\caption{\subref{fig:peakrm}: The Faraday depth at the peak PI in the region of Sh2-27 in the range $-3<\phi<+3\,$\radms. \subref{fig:peakmom}: The first moment of the Faraday spectrum computed in the range $-3<\phi<+3\,$\radms. White contours are \halpha\ intensity from \citet{Finkbeiner2003} at 30\,R. Black contours are of the peak PI (for all $\phi$) at 0.3\,K\,RMSF$^{-1}$. We label the five visible \hii\ regions, and their corresponding central stars (white stars), as in Figure~\ref{fig:peakpi}. We show the beam as a white circle in the lower-left corner. The range $-3<\phi<+3\,$\radms\ is used to select only the peak around 0\,\radms.}
\end{figure*}

We identify the large depolarized feature that extends to the right from Sh2-27 as a depolarization canal. Depolarization canals are a common feature of many polarization maps. These canals can occur from a variety of physical scenarios, but most commonly occur through one of two mechanisms \citep{Fletcher2006a,Fletcher2007}: either a strong gradient or discontinuity in Faraday depth across the sky, or depth depolarization along the LOS. Both of these mechanisms can produce depolarization which is the width of the telescope beam. In Figure~\ref{fig:peakrm} we show an image of the Faraday depth at the peak PI in the range $-3<\phi<+3\,$\radms. We select this restricted range in order to find the peak around 0\,\radms. The Faraday depth structure towards Sh2-27 is different to that along the feature. On Sh2-27 the peak $\phi$ is relatively smooth and constant ($\phi<0$). In contrast, there is a clear discontinuity in $\phi$ along the canal, as well as a gradient towards Galactic North. We confirm that these discontinuities are not artefacts of two peaks of similar heights by inspecting the first moment of the Faraday spectra in Figure~\ref{fig:peakmom}. This map shows the same discontinuities and gradients as the peak $\phi$ map, which indicates that these are true features of the Faraday depth structure. Areas with a discontinuity in $\phi$ show depolarization on the order of a beamwidth, which leads us to the conclusion that the feature is a depolarization canal. We note that the canal is slightly wider than the beamwidth, but this is explained by a combination of a discontinuity and a gradient in $\phi$. Both of these effects generate depolarization canals, and both appear in close proximity in the peak $\phi$ map. The depolarizing effects then blend into a wider canal. We conclude that this feature is distinct from Sh2-27 and we do not discuss it further.

\subsection{Faraday Spectra Towards Sh2-27}\label{sec:cubes}
\begin{figure*}
	\centering
	\includegraphics[width=\linewidth]{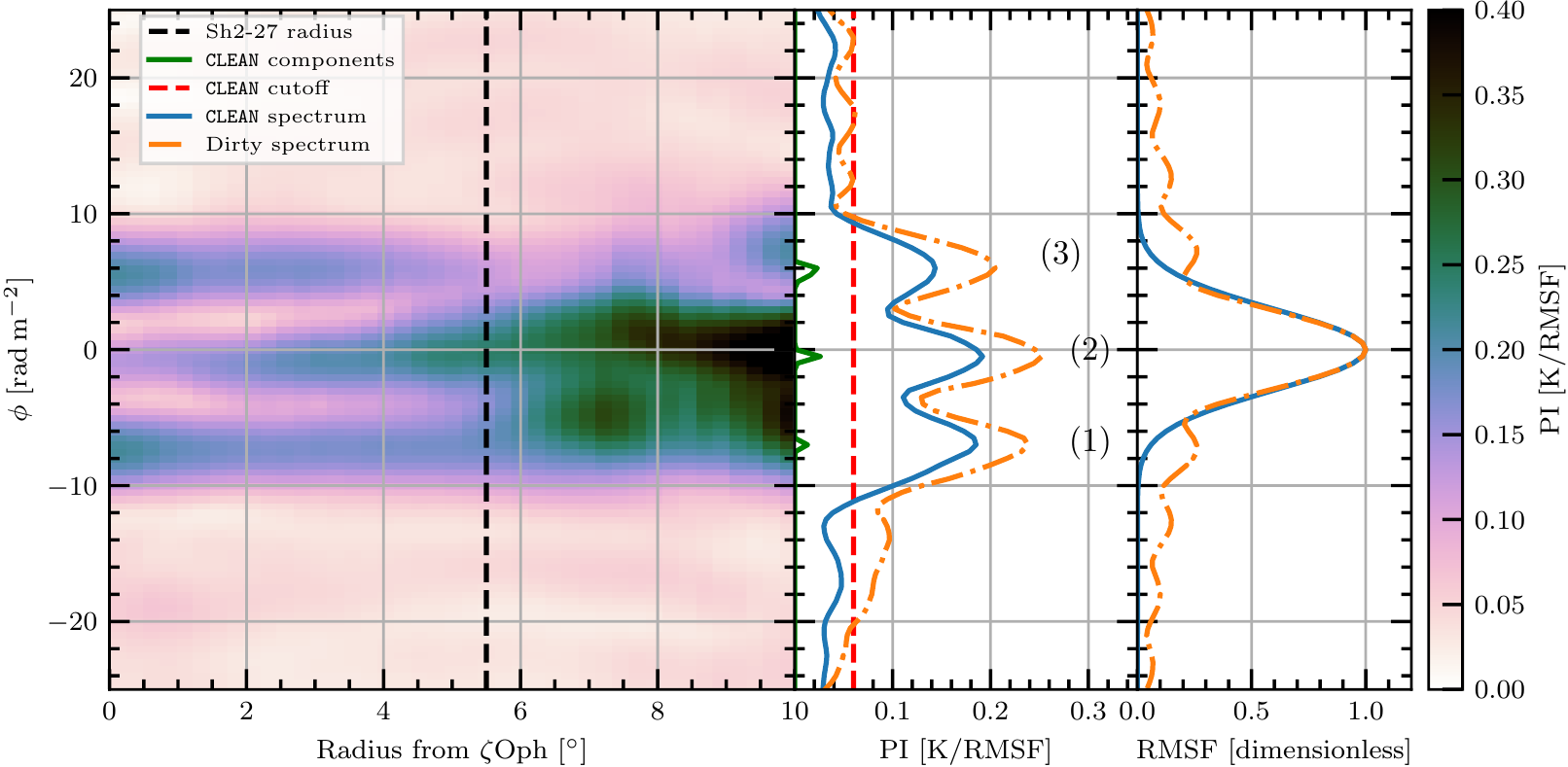}
	\caption{The Faraday depth structure towards Sh2-27. Left panel: Azimuthal averages of the Faraday spectrum as a function of radius from $\zeta$Oph. Middle panel: Median \texttt{CLEAN} and dirty Faraday spectrum, and \texttt{CLEAN} components, on the Sh2-27 region (as defined by \citet{Harvey-Smith2011a}). We also label the first, second, and third primary peaks. Right panel: Median dirty and \texttt{CLEAN} RMSF on the Sh2-27 region.}
	\label{fig:radprofile}
\end{figure*}

We find a consistent structure in the Faraday spectrum towards Sh2-27, shown in Figure~\ref{fig:radprofile}. In the left-hand panel of Figure~\ref{fig:radprofile} we show azimuthal averages (through a full rotation) of the Faraday spectrum in polarized intensity as a function of radius on the sky from $\zeta$Oph. For the region towards Sh2-27 we find a triple-peak structure, which is absent in the regions away from the \hii\ region. In the middle panel of Figure~\ref{fig:radprofile} we can see that each peak is well above our noise threshold and well fit by a single \verb|CLEAN| component. For comparison, we show the RMSF for the same region. It is clear that the triple-peak structure is not generated by sidelobes in the RMSF. The polarized intensity also increases significantly away from Sh2-27, correlating with the loss of the triple-peak structure. As the foreground structure is unlikely to correlate precisely with the boundary of Sh2-27, we conclude that the foreground structure we probe towards Sh2-27 is overwhelmed by higher intensity background emission in directions away from the depolarization wall.

To identify the Faraday depth of the peaks on Sh2-27 we first apply the peak-finding algorithm from \citet{Duarte2015} to find the Faraday-resolution-limited peaks in the azimuthally averaged spectra. We only search for peaks above our noise threshold of 60\,mK. From this we find the triple-peak structure extends radially for $5.5^{\circ}$ from $\zeta$Oph, which is almost exactly the radius of Sh2-27 in \halpha. We fit three Gaussians to the triple-peak region excluding structures below our noise threshold and obtain the means of the three peaks weighted by the inverse variance from the radial profile, 1: $-7.4\pm0.4\,$\radms, 2: $-0.8\pm0.4\,$\radms, and 3: $+6.2\pm0.4\,$\radms.

\section{Analysis}\label{sec:analysis}
When multiple peaks are present in a low-frequency Faraday spectrum two primary interpretations are possible: either the features are of separate origin, or the peaks arise from a Faraday thick medium which has been resolved out. We follow the method of \citet{VanEck2017} (hereafter CVE17) for separating these scenarios. We estimate the distance to the front of Sh2-27 using the distance to $\zeta$Oph. We use the parallax distance to this star from the \textit{Gaia} DR2 survey~\citep{GaiaCollaboration2016, GaiaCollaboration2018}, specifically the error-corrected distance estimates provided by \citet{Bailer-Jones2015}, $182\substack{+53 \\ -33}$\,pc. Taking the region to be a sphere centred on $\zeta$Oph with an angular radius of $5.5^{\circ}$ on the sky, we find the distance to the front of the region is $164\substack{+48 \\ -30}$\,pc.

\subsection{Faraday Thin Models Towards Sh2-27}\label{sec:modelism}
In this section we present a Faraday thin model of the foreground ISM towards Sh2-27, and show that it can accurately reproduce the observed Stokes $Q$ and $U$ spectra as a function of $\lambda^2$. We also consult additional data which can give information on the structure of the foreground column of ISM. 

\begin{figure*}
	\centering
	\includegraphics[width=\linewidth]{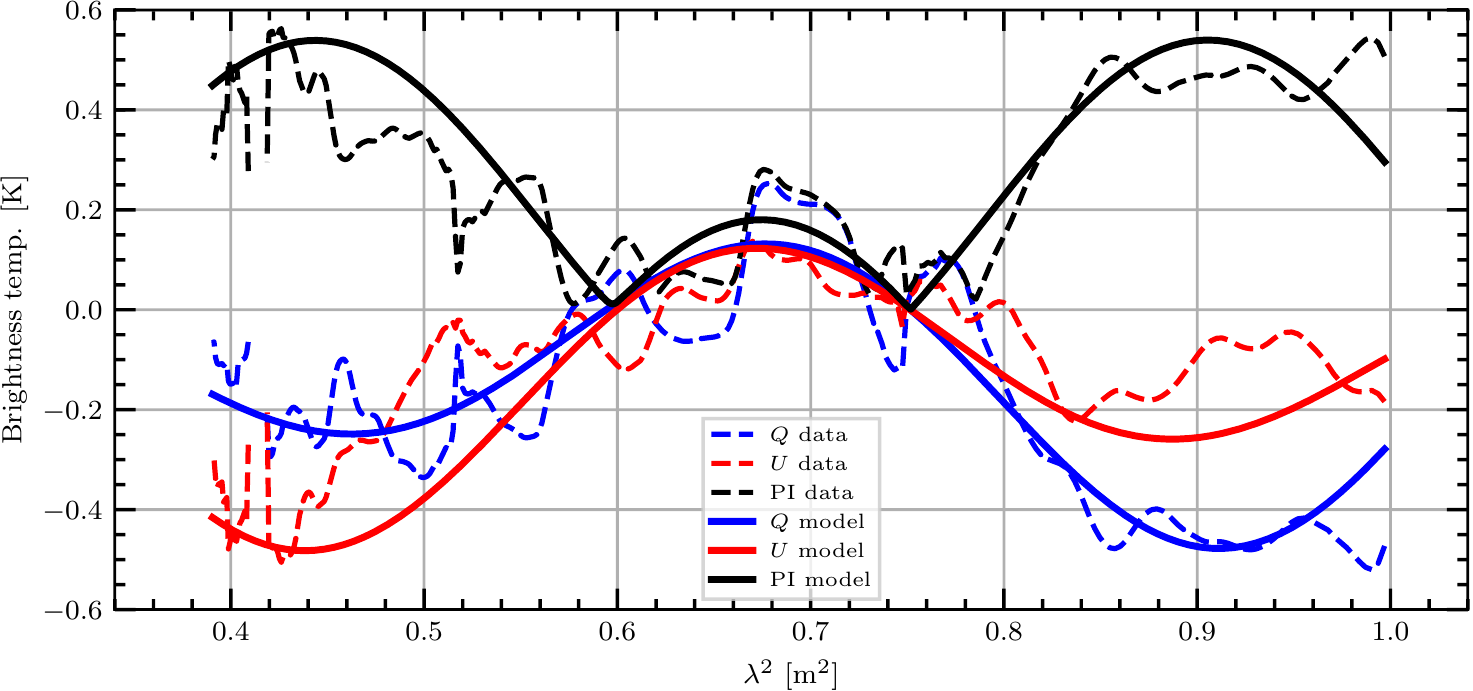}
	\caption{Faraday thin model spectra towards Sh2-27. Dashed lines: Average Stokes $Q$, $U$, and PI $\lambda^2$ spectra towards Sh-27 from GMIMS-LBS. Solid lines: Faraday thin model derived from the average Faraday spectrum.}
	\label{fig:thinmodel}
\end{figure*}

In general, the complex polarization of a Faraday thin component is given by:
\begin{equation}
	\mathcal{P}(\lambda^2) = \exp[2i(\chi_0+\phi_0\lambda^2)],
\end{equation}
where $\chi_0$ is the initial polarization angle of the emission and $\phi_0$ is the Faraday depth of the component. We obtain the de-rotated $\chi_0$ for peaks 1, 2, and 3 using:
\begin{equation}
	\chi_0 = \chi_1 - \phi_0\lambda^2_0\mod 180^{\circ},
\end{equation}
where $\chi_1$ is the polarization angle at the peak in the Faraday spectrum, and $\lambda^2_0$ is the de-rotated wavelength-squared as per \citet{Brentjens2005}. We construct model spectra as the sum of three Faraday thin components using the Faraday depth of each peak, their corresponding initial angles, and amplitudes of 0.18\,K.  We show both the average Stokes $Q$, $U$, and PI $\lambda$ spectrum on Sh2-27 and the Faraday thin model in Figure~\ref{fig:thinmodel}. We have not used any fitting routine, rather we have simply constructed the model from the average values we infer from the Faraday spectrum. 

There are two factors to consider as we construct a physical model of MIM along the LOS. We must consider where the polarized emission arises and determine to what degree the Faraday rotation occurs. We make this consideration under the constraint of the $\sim160\,$pc path to the front of Sh2-27. Meaning that we are analysing small, localised structures, with a size scale much less than a kiloparsec. We will first consider the sources of Faraday rotation before considering the source of polarized emission. A Faraday thin model does not necessarily exclude mixed emission and rotation, but for a model to be considered Faraday thin in our context the Faraday thickness should should not exceed the $\phi_{\text{max-scale}}$ of our observations.

The most likely contributors in the ISM to Faraday rotation of low frequency polarized emission are the cold and warm neutral medium (CNM and WNM), the warm ionised medium (WIM), and the hot ionised medium (HIM). There are no large molecular clouds towards Sh2-27, as indicated by the absence of obscuration of the \halpha\ emission from the \hii\ region. We can consider the amount of Faraday rotation each ISM phase is likely to contribute along the LOS, and quantify the path-length at which each phase will be resolved out of our observations. Here we take local electron densities of the various ISM phases from \citet{Ferriere2001} and \citet{Heiles2012}, and we assume a typical regular magnetic field value of $2\,\mu$G~\citep{Sun2007b} with no reversals. CVE17 conducted a similar analysis in the LOFAR band, finding that only emissions produced in the WNM would not be resolved out. We summarise these results in Table~\ref{tab:rotate}, comparing the survey characteristics from GMIMS-LBS and LOFAR. Since the $\phi_{\text{max-scale}}$ of GMIMS-LBS is nearly eight times that of LOFAR, our survey is much less susceptible to resolving out Faraday thick structures. We therefore cannot construct a similar model to CVE17, where interpretation of the polarized emission was tied to the absence of depolarization in the WNM. Instead, the features that we observe must be explained by enhancements in the MIM along the LOS.

\begin{table}
	\centering
	\caption{Faraday rotation properties for various ISM phases. Col.(1): The ISM phases. Col.(2): The local electron density of the ISM~\citep{Ferriere2001,Heiles2012}. Col.(3): The Faraday rotation per unit distance, assuming a $2\,\mu$G LOS magnetic field with no reversals. Col.(4) and (5): The depth along the LOS after which depth depolarization will filter out polarized emission for LOFAR and GMIMS, respectively.}

	\begin{tabular}{l c c c c c}
		\toprule
		\midrule
		Phase	& $n_e$			& Faraday 						& \multicolumn{2}{c}{Path length}				\\
		& [cm$^{-3}$]	& rotation  					& \multicolumn{2}{c}{[pc]}						\\
		&				& [\radms\,pc$^{-1}$]			& LOFAR 							& GMIMS-LBS	\\

		\midrule
		CNM		& 0.016			& 0.026 						& 42 								& 310		\\
		WNM		& 0.0007		& 0.0011 						& 1000								& 7300		\\
		WIM 	& 0.25			& 0.41 							& 2.7 								& 20		\\
		WPIM	& 0.1			& 0.16 							& 6.9 								& 50		\\
		HIM		& 0.0034		& 0.006 						& 200 								& 1400		\\
		\bottomrule
	\end{tabular}
	\label{tab:rotate}
\end{table}

The different ISM phases along the LOS will each contribute differently to the Faraday rotation of synchrotron emission, due to their different magneto-ionic properties. The Local Bubble consists of a hot ionised medium (HIM), at $n_e = 0.005\,$cm$^{-3}$~\citep{Cordes2002,Shelton2009}, filling a volume around the Sun. Synchrotron emissions produced inside the Local Bubble should create a peak in the Faraday spectrum around $0\,$\radms, as emission produced close to the Sun should experience minimal Faraday rotation. Our peak 2 is consistent with 0\,\radms\ at $2\sigma$. We therefore interpret peak 2 as emission that is produced within the Local Bubble. At $1\sigma$ of confidence, we observe $-0.8\pm0.4$\radms\ of Faraday rotation through this volume.

Faraday rotation in the Local Bubble also affects the features which arise behind it; that is, we must subtract the $-0.8$\,\radms\ contribution from peaks 1 and 3. Applying this moves peaks 1 and 3 to $-6.6\pm0.6\,$\radms\ and $+7.1\pm0.6$\,\radms, respectively. We can constrain what is producing these features by analysing how LOS components of the ISM are contributing to Faraday rotation. Taking our values from Table~\ref{tab:rotate}, assuming these phases are contributing $\sim7\,$\radms\ of Faraday rotation would require a path-length of about 270\,pc, 6\,kpc, 17\,pc, 40\,pc, and 1.2\,kpc respectively.

Because of the short path-length ($164\substack{+48 \\ -30}$\,pc) to the front of Sh2-27, the only possible candidates are the CNM, WIM, WPIM. Neutral gas is typically traced using \hi\ observations. We inspect the \hi\ emission in the region of Sh2-27 from HI4PI~\citep{BenBekhti2016}. Due to the proximity of Sh2-27 to the Sun, \hi\ emissions produced in this region crowd around 0\,km/s, making kinematic distances unreliable. We do find indications of \hi\ self-absorption, however, in the \hi\ spectra towards the \hii\ region, which indicates the presence of cold atomic gas. We are therefore motivated to look to the STILISM project \citep{Lallement2014,Capitanio2017,Lallement2018}, which traces the CNM and provides the LOS distances to these neutral structures.

\begin{figure*}
	\centering
	\begin{subfigure}[b]{0.49\textwidth}
		\includegraphics[width=\textwidth]{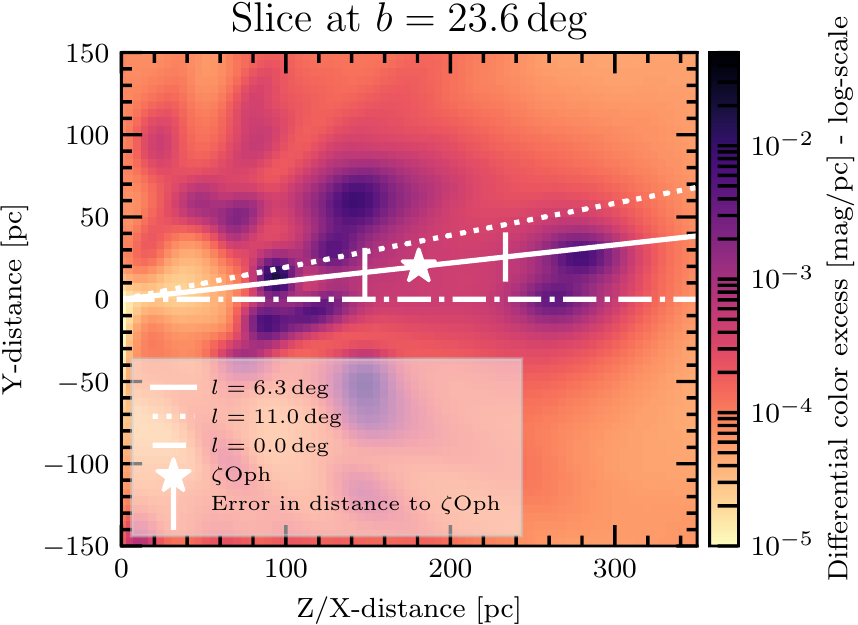}
		\subcaption{}
		\label{fig:dustslicelon}
	\end{subfigure}
	~ 
	\begin{subfigure}[b]{0.49\textwidth}
		\includegraphics[width=\linewidth]{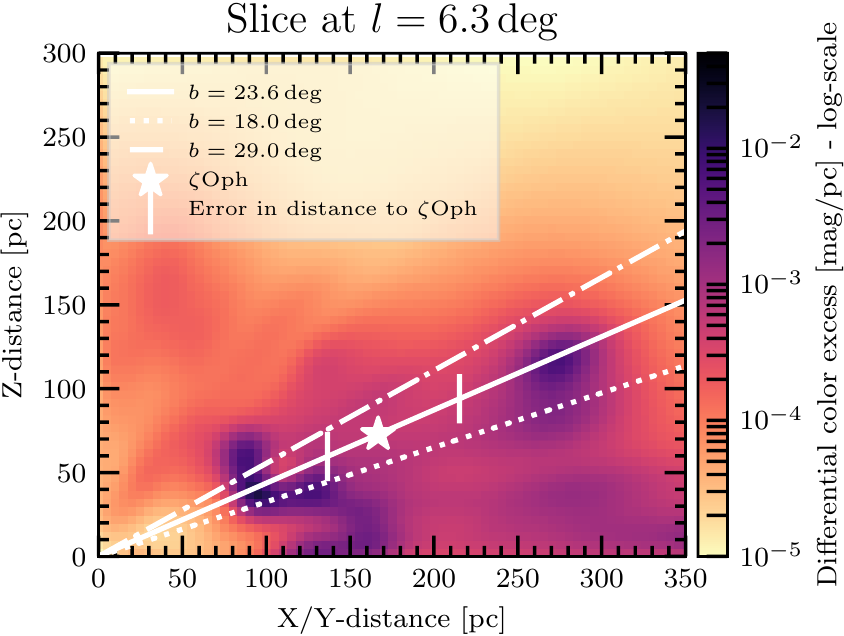}
		\subcaption{}
		\label{fig:dustslice}
	\end{subfigure}
	~ 
	\begin{subfigure}[b]{0.49\textwidth}
		\includegraphics[width=\linewidth]{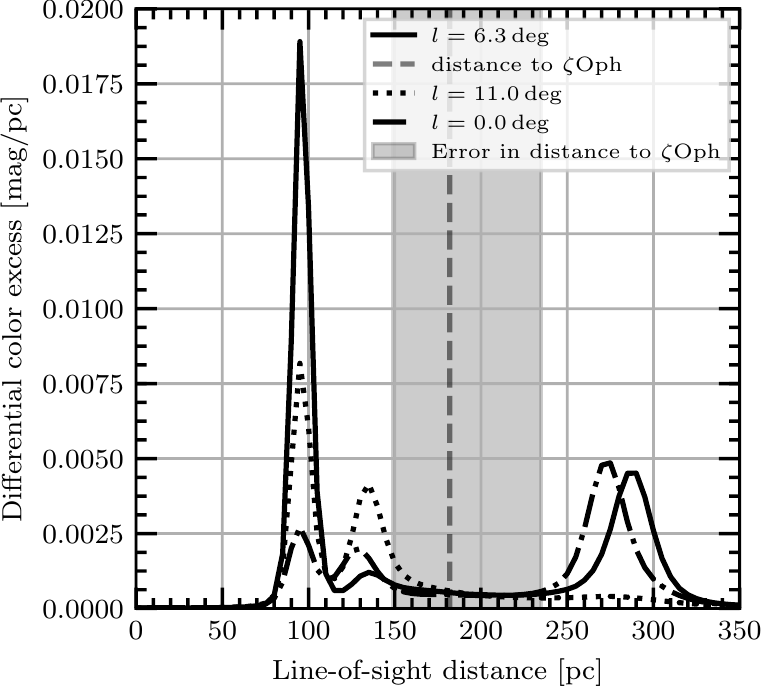}
		\subcaption{}
		\label{fig:dustprofilelon}
	\end{subfigure}
	~
	\begin{subfigure}[b]{0.49\textwidth}
		\includegraphics[width=\linewidth]{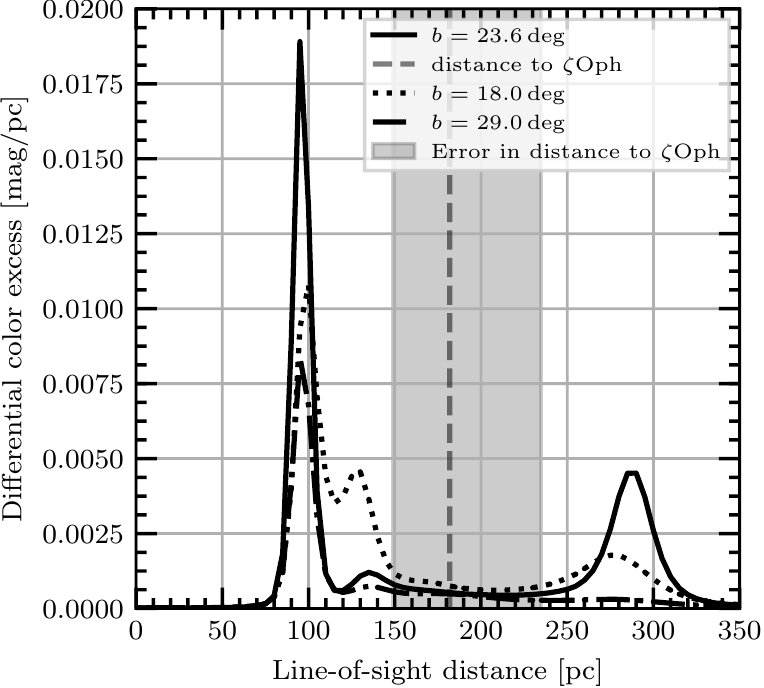}
		\subcaption{}
		\label{fig:dustprofile}
	\end{subfigure}
	\caption{Three-dimensional dust structure towards Sh2-27 from STILISM~\citep{Lallement2014,Capitanio2017,Lallement2018}. In all panels the solid line shows the LOS through the position of $\zeta$Oph, and the dashed lines are LOS through the outer bounds of the \hii\ regions. \subref{fig:dustslicelon} Slice through data cube at a constant latitude. \subref{fig:dustslice} Slice through data cube at a constant longitude. \subref{fig:dustprofilelon} and \subref{fig:dustprofile} show the LOS profiles for panels \subref{fig:dustslicelon} and \subref{fig:dustslice}, respectively.}
	\label{fig:slices}
\end{figure*}

We show a series of slices through the STILISM cube in Figure~\ref{fig:slices}. The Local Bubble appears as a void surrounding the Sun in these data. We find that the distance to edge of the Local Bubble is 80\,pc in the direction of Sh2-27. Taking an electron density of 0.005\,cm$^{-3}$ we derive a magnetic field strength of $-2.5\pm1.2\,\mu$G in the Local Bubble, aligned away from the Sun.

\begin{figure*}
	\centering
	\begin{subfigure}[b]{0.49\textwidth}
		\includegraphics[width=\textwidth]{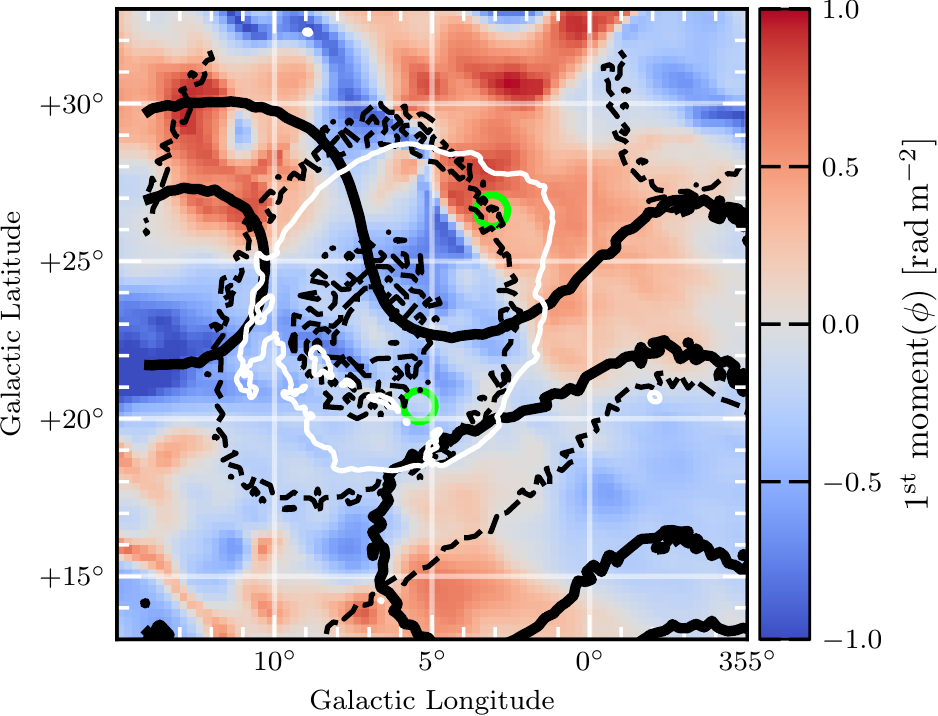}
		\caption{}
		\label{fig:clouds}%
	\end{subfigure}
	~ 
	\begin{subfigure}[b]{0.4\textwidth}
		\includegraphics[width=\textwidth]{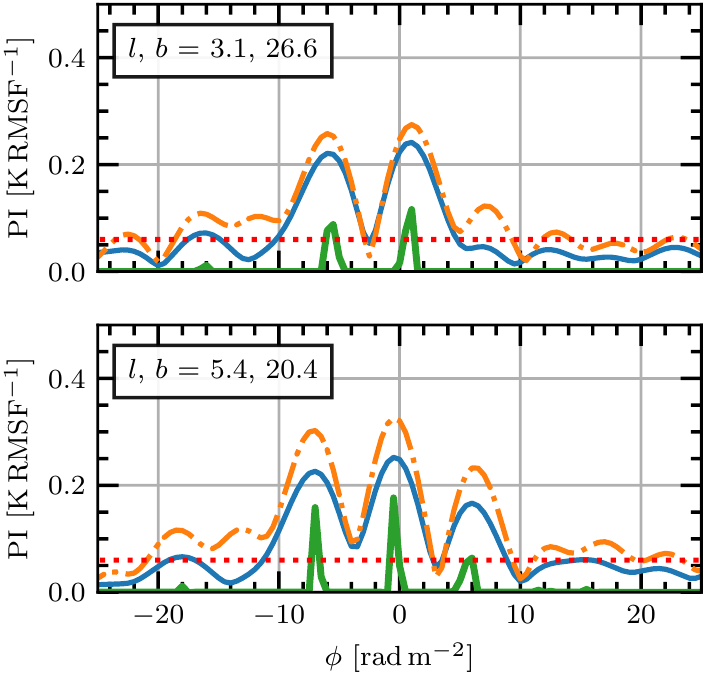}
		\caption{}
		\label{fig:spectra}%
	\end{subfigure}
	\caption{\subref{fig:clouds}: The first moment map of the Faraday spectrum (as in Figure~\ref{fig:peakmom}). White contours are \halpha\ intensity from \citet{Finkbeiner2003} at 30\,R. Black, dashed contours show STILISM dust reddening at 90\,pc, corresponding to the near neutral cloud. Black, solid contours show STILISM dust reddening at 135\,pc, corresponding to the far neutral cloud. Green circles show the positions for the Faraday spectra in the right-hand panel. \subref{fig:spectra}: Faraday spectra for two lines-of-sight towards Sh2-27. The upper panel shows a LOS which intersects with only the near cloud. The lower panel shows a LOS which intersects both neutral clouds. }
\end{figure*}

The location of Sh2-27 correlates with a region of relatively lower dust content in STILISM, as expected around an \hii\ region, compared to neutral clouds. Between the front of Sh2-27 and the edge of the Local Bubble two dust features appear. These regions occur at $\sim95$\,pc and $\sim135$\,pc and are each $\sim30$\,pc deep along the LOS. The distance error from the reddening inversion in this area is $\sim11\,$pc. We provide the spatial coverage of these clouds in the contours of Figure~\ref{fig:clouds}. The near cloud covers the entire region towards Sh2-27, whilst the far cloud only covers the lower-left portion of the region. Comparing the Faraday spectra between these areas we find that the triple-peak structure changes to a double-peak in the upper-right portion of the region, as shown in Figure~\ref{fig:spectra}. We see that there is neutral material in front of Sh2-27, and its location correlates with the Faraday spectra, so we can explain the Faraday properties of the foreground column without any WIM or WPIM along the line of sight. The magnetic fields need to be more intense, however, than the $\sim2\,\mu$G we assumed previously.

In higher density regions of the ISM magnetic fields become compressed \citep{Crutcher2010} and highly ordered, even in a relatively neutral medium \citep{Clark2014, Kalberla2017, Gazol2018, Tritsis2018}. The dust features towards Sh2-27 are composed of CNM, and thus are a higher density region of neutral ISM. We can estimate the density in these clouds using a dust-to-gas ratio. \citet{Liszt2014} find a ratio of \hi\ column density ($ N(\text{\hi}) $) to dust reddening magnitude ($E(B-V)$) of $N(\text{\hi}) = 8.3\times10^{21}\,\text{cm}^{-2}\,E(B-V)$ for for $|b|>20^{\circ}$ and $E(B-V)\lesssim0.1\,$mag. This corresponds to a number density ($n(\text{\hi})$) to differential colour excess ratio of $\sim2700\,\text{cm}^{-3}/(\text{mag}\,\text{pc}^{-1})$. For the two foreground clouds, we find a total number density of $n_\text{tot}\sim50\,$cm$^{-3}$ and $\sim12\,$cm$^{-3}$, which is consistent with typical values in the CNM~\citep{Ferriere2001}. Increased electron density and magnetic fields in the dust features are evidently providing increased Faraday rotation over the more tenuous inter-cloud medium.

\begin{figure*}
	\centering
	\def\svgwidth{1.1\textwidth}
	{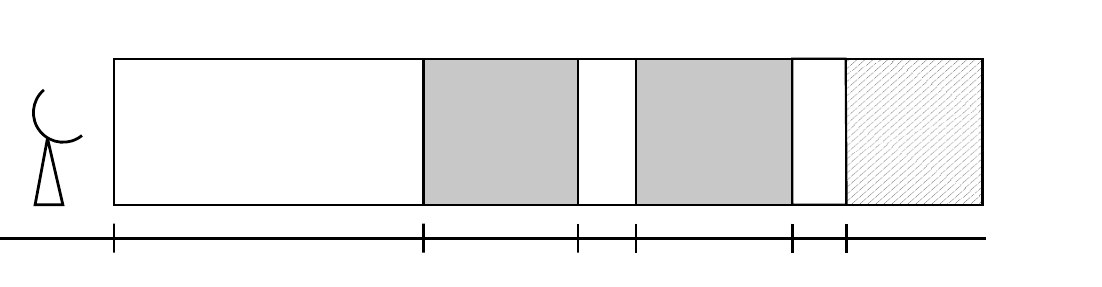}
	\caption{A cartoon of the magnetic field structure we observe along the LOS towards Sh2-27. We indicate the approximate distance to each feature along the bottom of the figure. We shade the two neutral clouds grey, indicating their increased density over other LOS components. The hatched region corresponds to the front of Sh2-27, behind which we receive no polarized emission. We give the values for the Faraday depths in each region. Arrows indicate the magnetic field direction in the Local Bubble and the two neutral clouds, as determined from our observations.}
	\label{fig:los}
\end{figure*}

The observed triple-peaked Faraday spectrum can be reproduced from a simple model of the magneto-ionic structure towards Sh2-27. We summarise this model of the MIM towards Sh2-27 in Figure~\ref{fig:los}. In this model we first assume a constant synchrotron emissivity ($\varepsilon$) along the entire LOS towards Sh2-27. We interpret peaks 1 and 3 to be associated with the dust features. Such peaks would be produced if both clouds have stronger Faraday rotation, with LOS magnetic fields of opposite directions and with the cloud further from the Sun having stronger LOS magnetic field than the closer one. This must be the case to produce two peaks. If the clouds had similar strength LOS magnetic fields, emission produced behind both clouds would be Faraday rotated by the closer cloud to $\sim0\,$\radms. Further, we are able to associate peak 3 with the far cloud from the change in the Faraday spectrum on and off the cloud. This means that peak 1 arises from the near cloud. To summarise, assuming a uniform $\varepsilon$, the triple peak structure can be created from the far cloud with a Faraday depth of $+13.7\pm0.8\,$\radms, the near cloud with Faraday depth of $-6.6\pm0.6\,$\radms, and a peak near 0\,\radms\ from the Local Bubble. Emission produced in the warm inter-cloud regions is not depolarized, but undergoes an increased amount of Faraday rotation in the neutral dust clouds.

\begin{figure*}
	\centering
	\begin{subfigure}[b]{0.48\textwidth}
		\includegraphics[width=\textwidth]{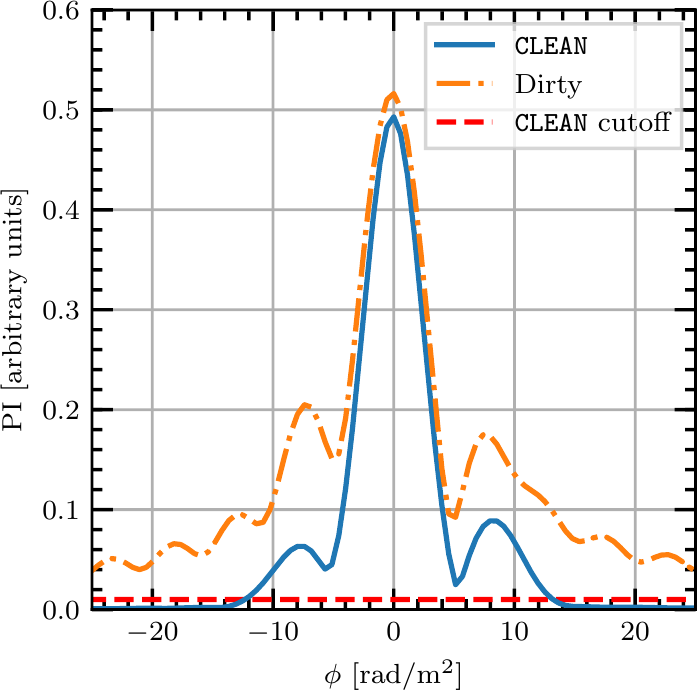}
		\caption{}
		\label{fig:simspec_uniform}%
	\end{subfigure}
	~ 
	\begin{subfigure}[b]{0.49\textwidth}
		\includegraphics[width=\textwidth]{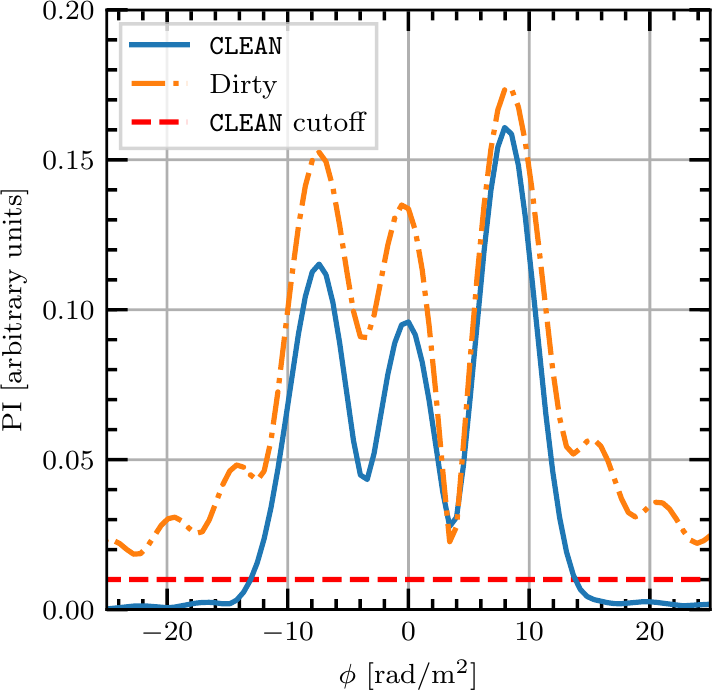}
		\caption{}
		\label{fig:simspec_lowbub}%
	\end{subfigure}
	\caption{Simulated Faraday spectra of our Faraday thin model. The LOS distribution of the MIM is identical for each model, with only the emissivity changing. \subref{fig:simspec_uniform} Uniform emissivity along the entire LOS. \subref{fig:simspec_lowbub} Emissivity in the Local Bubble reduced by $90\%$.}
	\label{fig:simspec}
\end{figure*}

We confirm the viability of the model by constructing a simple 1D numerical simulation of the Faraday rotation produced by this model. Into this model we input LOS values for $B_\parallel$, $n_e$, and pseudo-$\varepsilon$, scaling the total emission to 1 flux unit. From this we obtain Stokes $Q$ and $U$ in the GMIMS-LBS band and perform Faraday tomography. We show the resulting Faraday spectra in Figure~\ref{fig:simspec}. In this evaluation of the simulation, we take $B_\parallel$ in the near and far cloud to be $ -15\, \mu$G and $+30\,\mu$G, respectively, with the rest of the LOS having $2\,\mu$G. We find that the resulting Faraday spectrum is relatively insensitive to the sign of the intra-cloud and Local Bubble field directions. When we assume a uniform $\varepsilon$ we obtain a triple-peak spectrum which is dominated by the component near $0\,$\radms, as shown in Figure~\ref{fig:simspec_uniform}. This is likely because this is over estimating the contribution of emission from the Local Bubble. More realistically, the magnetic fields in the HIM of the Local Bubble are likely to be weak \citep{Hill2012,Hill2018a}, and therefore the $\varepsilon$ in this region should be reduced relative to the rest of the LOS. In Figure~\ref{fig:simspec_lowbub} we show the result of setting the $\varepsilon$ of the Local Bubble to be $10\%$ of the remaining $\varepsilon$. This produces three peaks of approximately equal height in the Faraday spectrum. It is possible that this same structure may arise from a more complicated LOS composition. In the absence of data to motivate such a model, this simulation demonstrates that our observed Faraday structure can be produced from a simple model.

We can also determine how tenable this model is by calculating the polarization fraction. To do this we must also estimate the total synchrotron intensity towards Sh2-27. As Sh2-27 is a depolarization wall, we need to only consider the synchrotron emission from in front of the region. \citet{Roger1999} measured the total intensity towards a number of \hii\ regions, including Sh2-27, at 22\,MHz and estimated the synchrotron emissivity. They find $\varepsilon=159\,$K/pc at 22\,MHz, but they note that the emissivity towards Sh2-27 was very high relative to other \hii\ regions, and that Sh2-27 might be not completely optically thick at 22\,MHz. We investigate whether this is the case using values from the literature. The opacity ($\tau$) of an \hii\ region at a particular frequency ($\nu$) is given by \citet{Mezger1967}:
\begin{equation}
	 \tau = {3.28 \times 10^{-7}}   \left( \frac{T_e}{10^4}\right) ^{-1.3} \left( \frac{\nu}{\text{[GHz]}}\right) ^{-2.1} \text{EM},
	 \label{eqn:tau}
\end{equation}
where $n_e\approx2$\,cm$^{-3}$ \citep{Wood2005}, $\text{EM} = 240\pm26$\,cm$^{-6}$\,pc \citep{Celnik1988} is the emission measure, and $T_e$ is the electron temperature. Taking $T_e=7000\,$K gives $\tau=0.38$ at 22\,MHz, meaning Sh2-27 is not optically thick. Using this opacity, we re-derive a foreground emissivity of $\varepsilon=37\substack{+23 \\ -15}$. More recently, \citet{Su2018} calculated the synchrotron emissivity towards many \hii\ regions at 76.2\,MHz using the Murchison Widefield Array (MWA). They find an average value of $1\pm0.5\,$K\,pc$^{-1}$ at 76.2\,MHz. Taking a spectral index of $\beta=-2.5$ (where $I\propto \nu^{\beta}$), the emissivity at the GMIMS-LBS mid-band frequency of 390\,MHz is $\varepsilon=0.017\pm0.008\,$K\,pc$^{-1}$. This value is also consistent with our recomputed value from \citet{Roger1999} assuming the same spectral index. Using the scaled emissivity from \citet{Su2018}, we estimate the total flux arising in front of Sh2-27 is $2.8\substack{+2.5 \\ -1.6}$\,K.

The use of depolarization walls is conceptually similar to using free-free absorption of Stokes $I$ by \hii\ regions. Similarly, we can determine the total received polarized emission towards Sh2-27. Using our Gaussian fit for the three Faraday thin components, we integrate the polarized intensity over the range of Faraday depths to determine the total polarized flux. From this we find a total polarized flux of $\sim0.4$\,K. Taking our previous estimate of the total intensity, this results in a polarization fraction of $12\substack{+16 \\ -6}$\,\%. Given that spatial variation in Faraday depth will cause significant beam depolarization, this fraction is relatively high. This value further supports our finding that the magnetic fields causing the observed Faraday rotation towards Sh2-27 have a highly ordered component.

Finally, we estimate the magnetic field strengths in the neutral clouds. We have determined that the far cloud has a Faraday depth of $\sim+14\,$\radms\, and the near cloud a Faraday depth of $\sim-7\,$\radms. From Equation~\ref{eqn:faradaydepth} we also need to estimate $n_e$, and the path-length through each region ($L$). We find no pulsars between the Sun and Sh2-27 in ATNF Pulsar Catalogue \citep{Manchester2005}\footnote{Catalogue version: 1.59, Accessed 26$^\text{th}$ of November 2018.}, and since Sh2-27 is the dominant \halpha\ emission source in this direction it is not possible to constrain the $n_e$ from these observations. As such, we present the LOS magnetic field strength as a function of the total number density ($n_\text{tot}$), the ionisation fraction ($X_e$), and $L$. We also estimate the strengths taking reasonable values from \citet{Ferriere2001} and our estimates above:
\begin{align*}
	B_{\parallel, \text{near}} \approx -15\,\mu\text{G} \left( \frac{n_\text{tot}}{20\,\text{cm}^{-3}}\right)  \left( \frac{X_e}{1\times10^{-3}}\right) \left(  \frac{L}{30\,\text{pc}}\right)\\
	B_{\parallel, \text{far}} \approx +30\,\mu\text{G} \left( \frac{n_\text{tot}}{20\,\text{cm}^{-3}}\right)  \left( \frac{X_e}{1\times10^{-3}}\right) \left(  \frac{L}{30\,\text{pc}}\right)
\end{align*}

\subsection{Faraday Thick Models Towards Sh2-27}\label{sec:modelcube}
We can also decide whether the Faraday structure towards Sh2-27 is Faraday thick using the CVE17 polarization flux method. After performing Faraday tomography, the PI spectra have units of K$/$RMSF. To obtain polarized flux, we must convert these units to K$/$(\radms). This conversion factor of \radms$/$RMSF is given by the integrated area ($A$) under the \verb|CLEAN| Gaussian RMSF. For the region towards Sh2-27 in GMIMS-LBS this factor is 7.3\,\radms$/$RMSF. Note, that for LOFAR observations CVE17 obtained a conversion factor of near unity, whereas the factor for GMIMS-LBS is nearly an order of magnitude higher.

\begin{figure}
	\centering
	\includegraphics[width=\linewidth]{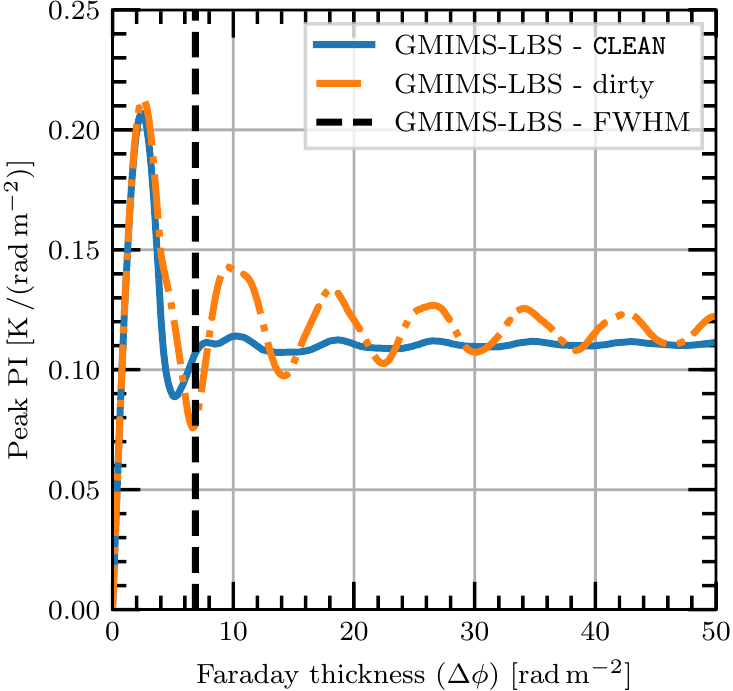}
	\caption{The depth depolarization of a Burn slab as a function of Faraday thickness, as observed by GMIMS-LBS. The peak PI is taken from a synthetic Faraday tomography observation of a Burn slab with a height 1\,K and a variable thickness. Blue, dash-dotted: Depolarization from dirty spectra. Orange, solid: Depolarization from CLEAN spectra. Black, dashed: The FWHM of the RMSF.}
	\label{fig:depthdepol}
\end{figure}

We can now model the depolarization of a Faraday thick medium in GMIMS-LBS. We model this as a `Burn slab' \citep{Burn1966}, the simplest Faraday thick model. In Faraday depth space a Burn slab is a tophat function, which corresponds to the following complex polarization in $\lambda^2$:
\begin{equation}
\mathcal{P}(\lambda^2) = \exp[2i(\chi_0+\phi_0\lambda^2)]\frac{\sin(\Delta\phi\lambda^2)}{\lambda^2},
\label{eqn:slab}
\end{equation}
where $\phi_0$ is the central Faraday depth of the slab, and $\Delta\phi$ is the width, or Faraday thickness, of the slab, and $\chi_0$ again is the initial polarization angle. This model has the additional advantage of resolving out the least as a function of Faraday thickness; that is, other Faraday thick models will be filtered out more strongly. We model observations using GMIMS-LBS by evaluating this complex polarization using $\lambda^2$ values observed by GMIMS-LBS, taking the height of the slab to be 1\,K, and then performing Faraday tomography on the resulting spectra. As the model is resolved out, the `observed' Faraday spectrum is split into two peaks which also reduce in magnitude. We show this reduction as a function of Faraday thickness \citep[matching Figure A.1. of][]{VanEck2017} in Figure~\ref{fig:depthdepol}. We note that this function is smooth compared to CVE17 because we have also applied \verb|RM-CLEAN| to our synthetic spectra (not doing so results in an oscillation due to interference between the sidelobes of the depolarized peaks). We find that if the Faraday thickness of the slab is greater than the FWHM of the RMSF, then the depolarization factor is about $11\%$. For a Faraday thickness less than that, the depolarization factor varies significantly, reaching a peak depolarization factor of about $21\%$ at 2.4\,\radms.

There are three possible thick models that could apply to our observations (1): either peaks 1 and 2 are edges of a thick slab, (2): peaks 2 and 3 are edges of a slab, or (3): peaks 1 and 3 are the edges of the slab. In each case the third peak would be provided by a Faraday thin component. We will only consider cases (1) and (2), as case (3) will result in greater missing flux. In both cases we cannot know which peak represents the leading edge of a slab \textit{a priori}. This condition, however, only sets the direction of the coherent magnetic field along the LOS, and does not affect the degree of missing flux. The Faraday thicknesses for models (1) and (2) are $6.6\pm0.6\,$\radms and $7.1\pm0.6\,$\radms, respectively. The heights of peaks 1, 2, and 3 are $0.185\pm0.002$\,K/RMSF, $0.190\pm0.005$\,K/RMSF, and $0.168\pm0.006$\,K/RMSF, respectively. For simplicity, we can consider both of these cases together as a slab of thickness $\sim7\,$\radms, and a depolarized peak of $\sim0.18$K/RMSF. Taking the conversion factor of 7.3\,\radms$/$RMSF gives the height of the depolarized peak as $\sim0.024\,$K/(\radms). A Faraday thickness of $\sim7\,$\radms\ will correspond to a depolarization factor of $\sim11\%$, and therefore the height of the slab will be $\sim0.23$\,K/(\radms). Integrating across the slab results in a polarized flux of $\sim1.6$\,K. From our estimate above, a Faraday thin component would provide about $0.1\,$K of flux.

For the $\varepsilon$ we calculate above, the polarization fraction would therefore be $62\substack{+81 \\ -29}$\,\%. For comparison, the maximum theoretical polarization fraction for synchrotron emission is 75\% \citep{Rybicki1986}, but this will only occur when the magnetic field generating the synchrotron emission is perfectly uniform. Such high values are highly unlikely to arise in the diffuse ISM.

We also evaluate the $\lambda^2$ spectra for each Burn slab model in a similar manner to the Faraday thin case. We show the resulting spectra in Section~\ref{sec:appthick}. None of these models recreate the average spectra well, especially in comparison to the thin model. From both this finding, and our analysis of the polarized flux from a Burn slab model, we conclude that a Faraday thick model is unlikely to apply here. 

\section{Discussion}\label{sec:discuss}
Faraday tomography is a powerful method for probing the MIM of the Galactic ISM. Faraday depth, however, can vary in a non-monotonic fashion along the LOS and mapping structure in the Faraday dispersion function is therefore difficult. The use of depolarization to constrain distances to polarized features has been applied in many diffuse polarization surveys \citep[e.g.][]{Wolleben2010,Hill2017}. We have shown that at low frequencies this analysis can be extended. If a depolarization feature can be identified as a depolarization wall then any observed polarized emission can be constrained to the region along the LOS in front of the feature. In GMIMS-LBS we are sensitive to large angular scales, but our large beam also constrains us to this type of analysis only on large depolarization regions. Additionally, the current spatial density of extragalactic RMs \citep[e.g.][]{Taylor2009} is $\sim1\,$RM/deg$^2$, which also restricts the analysis of beam depolarization Future polarized surveys, such as POSSUM~\citep{Gaensler2010} from the Australian SKA Pathfinder (ASKAP), aim to deliver $\sim100$\,RM/deg$^2$. With such data, the type of analysis we present here can be extended to higher angular resolution with observations from aperture synthesis telescopes. Furthermore, distances to \hii\ regions are being well constrained by the \hii\ Region Discovery Surveys \citep[HRDS, SHRDS][]{Bania2010,Brown2017b}.

Understanding of the density-magnetic field relationship in the ISM is of great importance to many processes. Recent observations \citep[e.g.][]{Wolleben2010,Clark2014, Kalberla2017,Tritsis2018} and numerical simulations \citep[e.g.][]{Gazol2018} have shown that even in the diffuse ISM magnetic fields can be compressed and ordered. Our observations are highly compatible with this picture, and our model of the ISM towards Sh2-27 shows that magnetic fields have become ordered and magnified in nearby dust clouds. \citet{Crutcher2010} show that in densities associated with the CNM, magnetic fields are measured be on the order of $5\mu\,$G, but can be as high as $10$--$20\,\mu$G. \citet{Wolleben2010} use Faraday tomography to measure the magnetic field in large, nearby \hi\ shell. They determine a LOS field strength of $20$--$34\,\mu$G. \citet{Clark2014} estimate a total magnetic field strength in the Riegel-Crutcher \hi\ cloud of $10$--$50\,\mu$G, using a Chandrasekhar-Fermi-like method. \citet{McClureGriffiths2006} previously constrained that the total magnetic field in the Riegel-Crutcher cloud should be at least $30\,\mu$G. \citet{Tritsis2018} analyse a similar region in Ursa Major, finding a total magnetic field strength of $10$--$20\,\mu$G. Our magnetic field estimates are broadly consistent with these measurements. We note however, that each of these cases represents an atypical cloud, as compared with \citet{Crutcher2010} results for the same density. Further investigation of the clouds we find towards Sh2-27 is required to understand whether such a special case, such as compression within a shell wall, occurs here.

\section{Summary and Conclusion}\label{sec:summary}
In this paper we have made use of the highly sensitive GMIMS-LBS observations to probe the magneto-ionic structure of the nearby ISM. We achieve this by identifying the nearby \hii\ region Sh2-27 as a depolarization wall. The magneto-ionic properties of Sh2-27, as revealed by extragalactic RMs, prevent polarized emissions produced behind the region at 300--480\,MHz from propagating through it. We are then able to perform Faraday tomography on the observed polarized emission knowing that the structure we observe must originate between the Sun and the front of Sh2-27, a path length of only 160\,pc.

We find a consistent triple-peaked structure in the Faraday spectrum in the region towards Sh2-27. We conclude that the structure is highly unlikely to arise from a resolved out Faraday thick source, but rather should be caused by magneto-ionic enhancements along the LOS. We draw this conclusion from both consideration of the polarized flux and by modelling Faraday thick and thin spectra. We find that only the thin model reproduces the observations well.

Using three-dimensional ISM maps we identify two neutral features in front of Sh2-27 as well as the ionised region of the Local Bubble. The Local Bubble extends for 80\,pc in the direction of Sh2-27, and the two clouds lie in the remaining space in front of Sh2-27 and are each about 30\,pc thick. Given the constraint on the LOS structure we also find that the observed Faraday structure cannot arise from a tenuous ionised region. Rather, the structure must arise from magneto-ionic enhancements. We are able to associate the three peaks in our Faraday spectrum with the two neutral clouds and the Local Bubble. We confirm the viability of this model using both a simple 1D simulation, and an analysis of the polarized flux. Following this, we find a Faraday depth in the local bubble of $-0.8\pm0.4$\radms, meaning that magnetic field is aligned away from the Sun in this direction. Assuming that this Faraday rotation occurs uniformly throughout the Local Bubble, this Faraday depth corresponds to a LOS magnetic field strength of $-2.5\pm1.2\,\mu$G. In the near and far clouds we obtain Faraday depths $-6.6\pm0.6\,$\radms\ and  $+13.7\pm0.8\,$\radms, respectively. These Faraday depths correspond to LOS magnetic fields of opposite alignment in each cloud.

Here we have considered only a small region in the GMIMS-LBS. We chose this region as the morphological correlation between the polarization structure and the \hii\ region Sh2-27 is immediately apparent.  We have shown that interpretation of features in these data requires careful analysis and combination with extragalactic polarization observations and additional tracers of the ISM. We have shown that GMIMS observations are highly complementary to newly released survey data such as \textit{Gaia} and will be of great use for interpretation of results from the upcoming MWA and ASKAP surveys.

\section*{Acknowledgements}
The authors wish to thank JinLin Han for his constructive input.

AT acknowledges the support of the Australian Government Research Training Program (RTP) Scholarship. N. M. M.-G. acknowledges the support of the Australian Research Council through grant FT150100024. C.~F.~acknowledges funding provided by the Australian Research Council (Discovery Project DP170100603 and Future Fellowship FT180100495), and the Australia-Germany Joint Research Cooperation Scheme (UA-DAAD).

The Parkes Radio Telescope is part of the Australia Telescope National Facility which is funded by the Commonwealth of Australia for operation as a national facility managed by CSIRO. This work has made use of data from the European Space Agency (ESA) mission
{\it Gaia} (\url{https://www.cosmos.esa.int/gaia}), processed by the {\it Gaia} Data Processing and Analysis Consortium (DPAC, \url{https://www.cosmos.esa.int/web/gaia/dpac/consortium}). Funding for the DPAC has been provided by national institutions, in particular the institutions participating in the {\it Gaia} Multilateral Agreement. We further acknowledge high-performance computing resources provided by the Australian National Computational Infrastructure (grant~ek9) in the framework of the National Computational Merit Allocation Scheme and the ANU Allocation Scheme.
This research made use of Astropy,\footnote{http://www.astropy.org} a community-developed core Python package for Astronomy \citep{Robitaille2013, Price-Whelan2018}. This research made use of APLpy, an open-source plotting package for Python \citep{Robitaille2012}. We have made use of the `cubehelix' colour-scheme~\citep{Green2011}.




\bibliographystyle{mnras}
\bibliography{correct} 

\newcommand{\noop}[1]{}
\begin{thebibliography}{}
\makeatletter
\relax
\def\mn@urlcharsother{\let\do\@makeother \do\$\do\&\do\#\do\^\do\_\do\%\do\~}
\def\mn@doi{\begingroup\mn@urlcharsother \@ifnextchar [ {\mn@doi@}
  {\mn@doi@[]}}
\def\mn@doi@[#1]#2{\def\@tempa{#1}\ifx\@tempa\@empty \href
  {http://dx.doi.org/#2} {doi:#2}\else \href {http://dx.doi.org/#2} {#1}\fi
  \endgroup}
\def\mn@eprint#1#2{\mn@eprint@#1:#2::\@nil}
\def\mn@eprint@arXiv#1{\href {http://arxiv.org/abs/#1} {{\tt arXiv:#1}}}
\def\mn@eprint@dblp#1{\href {http://dblp.uni-trier.de/rec/bibtex/#1.xml}
  {dblp:#1}}
\def\mn@eprint@#1:#2:#3:#4\@nil{\def\@tempa {#1}\def\@tempb {#2}\def\@tempc
  {#3}\ifx \@tempc \@empty \let \@tempc \@tempb \let \@tempb \@tempa \fi \ifx
  \@tempb \@empty \def\@tempb {arXiv}\fi \@ifundefined
  {mn@eprint@\@tempb}{\@tempb:\@tempc}{\expandafter \expandafter \csname
  mn@eprint@\@tempb\endcsname \expandafter{\@tempc}}}

\bibitem[\protect\citeauthoryear{Bailer-Jones}{Bailer-Jones}{2015}]{Bailer-Jones2015}
Bailer-Jones C. A.~L.,  2015, \mn@doi [Publications of the Astronomical Society
  of the Pacific] {10.1086/683116}, 127, 994

\bibitem[\protect\citeauthoryear{{Bania}, {Anderson}, {Balser}  \&
  {Rood}}{{Bania} et~al.}{2010}]{Bania2010}
{Bania} T.~M.,  {Anderson} L.~D.,  {Balser} D.~S.,   {Rood} R.~T.,  2010,
  \mn@doi [\apj] {10.1088/2041-8205/718/2/L106}, \href
  {https://ui.adsabs.harvard.edu/#abs/2010ApJ...718L.106B} {718, L106}

\bibitem[\protect\citeauthoryear{Beck}{Beck}{2016}]{Beck2016}
Beck R.,  2016, \mn@doi [The Astronomy and Astrophysics Review]
  {10.1007/s00159-015-0084-4}, 24, 4

\bibitem[\protect\citeauthoryear{Beck \& Wielebinski}{Beck \&
  Wielebinski}{2013}]{Beck2013}
Beck R.,  Wielebinski R.,  2013, in , Vol.~68, Planets, Stars and Stellar
  Systems.
Springer Netherlands, Dordrecht, pp 641--723 (\mn@eprint {arXiv} {1302.5663}),
  \mn@doi{10.1007/978-94-007-5612-0_13}

\bibitem[\protect\citeauthoryear{{Ben Bekhti} et~al.,}{{Ben Bekhti}
  et~al.}{2016}]{BenBekhti2016}
{Ben Bekhti} N.,  et~al., 2016, \mn@doi [\aap] {10.1051/0004-6361/201629178},
  594, A116

\bibitem[\protect\citeauthoryear{Brentjens \& de Bruyn}{Brentjens \&
  de~Bruyn}{2005}]{Brentjens2005}
Brentjens M.~A.,  de Bruyn A.~G.,  2005, \mn@doi [\aap]
  {10.1051/0004-6361:20052990}, 441, 1217

\bibitem[\protect\citeauthoryear{{Brown} et~al.,}{{Brown}
  et~al.}{2017}]{Brown2017b}
{Brown} C.,  et~al., 2017, \mn@doi [\aj] {10.3847/1538-3881/aa71a7}, \href
  {https://ui.adsabs.harvard.edu/#abs/2017AJ....154...23B} {154, 23}

\bibitem[\protect\citeauthoryear{Burn}{Burn}{1966}]{Burn1966}
Burn B.~J.,  1966, \mn@doi [\mnras] {10.1093/mnras/133.1.67}, 133, 67

\bibitem[\protect\citeauthoryear{Capitanio, Lallement, Vergely, Elyajouri  \&
  Monreal-Ibero}{Capitanio et~al.}{2017}]{Capitanio2017}
Capitanio L.,  Lallement R.,  Vergely J.~L.,  Elyajouri M.,   Monreal-Ibero A.,
   2017, \mn@doi [\aap] {10.1051/0004-6361/201730831}, 606, A65

\bibitem[\protect\citeauthoryear{{Carretti} et~al.,}{{Carretti}
  et~al.}{2019}]{Carretti2019}
{Carretti} E.,  et~al., 2019, arXiv e-prints, \href
  {https://ui.adsabs.harvard.edu/abs/2019arXiv190309420C} {p. arXiv:1903.09420}

\bibitem[\protect\citeauthoryear{{Celnik} \& {Weiland}}{{Celnik} \&
  {Weiland}}{1988}]{Celnik1988}
{Celnik} W.~E.,  {Weiland} H.,  1988, \aap, \href
  {https://ui.adsabs.harvard.edu/\#abs/1988A&A...192..316C} {192, 316}

\bibitem[\protect\citeauthoryear{Clark, Peek  \& Putman}{Clark
  et~al.}{2014}]{Clark2014}
Clark S.~E.,  Peek J. E.~G.,   Putman M.~E.,  2014, \mn@doi [\apj]
  {10.1088/0004-637X/789/1/82}, 789, 82

\bibitem[\protect\citeauthoryear{Condon, Cotton, Greisen, Yin, Perley, Taylor
  \& Broderick}{Condon et~al.}{1998}]{Condon1998}
Condon J.~J.,  Cotton W.~D.,  Greisen E.~W.,  Yin Q.~F.,  Perley R.~A.,  Taylor
  G.~B.,   Broderick J.~J.,  1998, \mn@doi [\aj] {10.1086/300337}, 115, 1693

\bibitem[\protect\citeauthoryear{Cordes \& Lazio}{Cordes \&
  Lazio}{2002}]{Cordes2002}
Cordes J.~M.,  Lazio T. J.~W.,  2002

\bibitem[\protect\citeauthoryear{Crutcher, Wandelt, Heiles, Falgarone  \&
  Troland}{Crutcher et~al.}{2010}]{Crutcher2010}
Crutcher R.~M.,  Wandelt B.,  Heiles C.,  Falgarone E.,   Troland T.~H.,  2010,
  \mn@doi [\apj] {10.1088/0004-637X/725/1/466}, 725, 466

\bibitem[\protect\citeauthoryear{{Dennison}, {Simonetti}  \&
  {Topasna}}{{Dennison} et~al.}{1998}]{Dennison1998}
{Dennison} B.,  {Simonetti} J.~H.,   {Topasna} G.~A.,  1998, \mn@doi
  [Publications of the Astronomical Society of Australia] {10.1071/AS98147},
  \href {https://ui.adsabs.harvard.edu/abs/1998PASA...15..147D} {15, 147}

\bibitem[\protect\citeauthoryear{{Dickey} et~al.,}{{Dickey}
  et~al.}{2019}]{Dickey2018}
{Dickey} J.~M.,  et~al., 2019, \mn@doi [\apj] {10.3847/1538-4357/aaf85f}, \href
  {https://ui.adsabs.harvard.edu/\#abs/2019ApJ...871..106D} {871, 106}

\bibitem[\protect\citeauthoryear{Duarte}{Duarte}{2015}]{Duarte2015}
Duarte M.,  2015, Notes on Scientific Computing for Biomechanics and Motor
  Control, \url{https://github.com/demotu/BMC}

\bibitem[\protect\citeauthoryear{Federrath}{Federrath}{2015}]{Federrath2015}
Federrath C.,  2015, \mn@doi [\mnras] {10.1093/mnras/stv941}, 450, 4035

\bibitem[\protect\citeauthoryear{Federrath \& Klessen}{Federrath \&
  Klessen}{2012}]{Federrath2012}
Federrath C.,  Klessen R.~S.,  2012, \mn@doi [\apj]
  {10.1088/0004-637X/761/2/156}, 761, 156

\bibitem[\protect\citeauthoryear{Federrath, Schr{\"{o}}n, Banerjee  \&
  Klessen}{Federrath et~al.}{2014}]{Federrath2014}
Federrath C.,  Schr{\"{o}}n M.,  Banerjee R.,   Klessen R.~S.,  2014, \mn@doi
  [\apj] {10.1088/0004-637X/790/2/128}, 790, 128

\bibitem[\protect\citeauthoryear{Ferri{\`{e}}re}{Ferri{\`{e}}re}{2001}]{Ferriere2001}
Ferri{\`{e}}re K.~M.,  2001, \mn@doi [Reviews of Modern Physics]
  {10.1103/RevModPhys.73.1031}, 73, 1031

\bibitem[\protect\citeauthoryear{Finkbeiner}{Finkbeiner}{2003}]{Finkbeiner2003}
Finkbeiner D.~P.,  2003, \mn@doi [\apjs] {10.1086/374411}, 146, 407

\bibitem[\protect\citeauthoryear{{Fletcher} \& {Shukurov}}{{Fletcher} \&
  {Shukurov}}{2006}]{Fletcher2006a}
{Fletcher} A.,  {Shukurov} A.,  2006, \mn@doi [\mnras]
  {10.1111/j.1745-3933.2006.00200.x}, \href
  {https://ui.adsabs.harvard.edu/\#abs/2006MNRAS.371L..21F} {371, L21}

\bibitem[\protect\citeauthoryear{Fletcher \& Shukurov}{Fletcher \&
  Shukurov}{2007}]{Fletcher2007}
Fletcher A.,  Shukurov A.,  2007, \mn@doi [EAS Publications Series]
  {10.1051/eas:2007008}, 23, 109

\bibitem[\protect\citeauthoryear{{Gaensler}, {Dickey}, {McClure-Griffiths},
  {Green}, {Wieringa}  \& {Haynes}}{{Gaensler} et~al.}{2001}]{Gaensler2001}
{Gaensler} B.~M.,  {Dickey} J.~M.,  {McClure-Griffiths} N.~M.,  {Green} A.~J.,
  {Wieringa} M.~H.,   {Haynes} R.~F.,  2001, \mn@doi [\apj] {10.1086/319468},
  \href {http://adsabs.harvard.edu/abs/2001ApJ...549..959G} {549, 959}

\bibitem[\protect\citeauthoryear{{Gaensler}, {Landecker}, {Taylor}  \& {POSSUM
  Collaboration}}{{Gaensler} et~al.}{2010}]{Gaensler2010}
{Gaensler} B.~M.,  {Landecker} T.~L.,  {Taylor} A.~R.,   {POSSUM Collaboration}
  2010, in American Astronomical Society Meeting Abstracts \#215. p. 470.13

\bibitem[\protect\citeauthoryear{{Gaia Collaboration} et~al.,}{{Gaia
  Collaboration} et~al.}{2016}]{GaiaCollaboration2016}
{Gaia Collaboration} et~al., 2016, \mn@doi [\aap]
  {10.1051/0004-6361/201629272}, 595, A1

\bibitem[\protect\citeauthoryear{{Gaia Collaboration} et~al.,}{{Gaia
  Collaboration} et~al.}{2018}]{GaiaCollaboration2018}
{Gaia Collaboration} et~al., 2018, \mn@doi [\aap]
  {10.1051/0004-6361/201833051}, 616, A1

\bibitem[\protect\citeauthoryear{Gaustad, McCullough, Rosing  \& {Van
  Buren}}{Gaustad et~al.}{2001}]{Gaustad2001}
Gaustad J.~E.,  McCullough P.~R.,  Rosing W.,   {Van Buren} D.,  2001, \mn@doi
  [Publications of the Astronomical Society of the Pacific] {10.1086/323969},
  113, 1326

\bibitem[\protect\citeauthoryear{Gazol \& Villagran}{Gazol \&
  Villagran}{2018}]{Gazol2018}
Gazol A.,  Villagran M.~A.,  2018, \mn@doi [\mnras] {10.1093/mnras/sty1041},
  478, 146

\bibitem[\protect\citeauthoryear{Green}{Green}{2011}]{Green2011}
Green D.~A.,  2011, Bull. Astr. Soc. India, pp 39--289

\bibitem[\protect\citeauthoryear{Green}{Green}{2014}]{Green2014}
Green D.~A.,  2014, Bull. Astr. Soc. India, 42, 47

\bibitem[\protect\citeauthoryear{Green et~al.,}{Green et~al.}{2018}]{Green2018}
Green G.~M.,  et~al., 2018, \mn@doi [\mnras] {10.1093/mnras/sty1008}, 478, 651

\bibitem[\protect\citeauthoryear{Haffner, Reynolds, Tufte, Madsen, Jaehnig  \&
  Percival}{Haffner et~al.}{2003}]{Haffner2003}
Haffner L.~M.,  Reynolds R.~J.,  Tufte S.~L.,  Madsen G.~J.,  Jaehnig K.~P.,
  Percival J.~W.,  2003, \mn@doi [\apjs] {10.1086/378850}, 149, 405

\bibitem[\protect\citeauthoryear{{Han}}{{Han}}{2017}]{Han2017}
{Han} J.~L.,  2017, \mn@doi [\araa] {10.1146/annurev-astro-091916-055221},
  \href {https://ui.adsabs.harvard.edu/abs/2017ARA&A..55..111H} {55, 111}

\bibitem[\protect\citeauthoryear{Harvey-Smith, Madsen  \&
  Gaensler}{Harvey-Smith et~al.}{2011}]{Harvey-Smith2011a}
Harvey-Smith L.,  Madsen G.~J.,   Gaensler B.~M.,  2011, \mn@doi [\apj]
  {10.1088/0004-637X/736/2/83}, 736, 83

\bibitem[\protect\citeauthoryear{Haverkorn, Katgert  \& de Bruyn}{Haverkorn
  et~al.}{2004}]{Haverkorn2004}
Haverkorn M.,  Katgert P.,   de Bruyn A.~G.,  2004, \mn@doi [\aap]
  {10.1051/0004-6361:200400051}, 427, 549

\bibitem[\protect\citeauthoryear{Heald, Braun  \& Edmonds}{Heald
  et~al.}{2009}]{Heald2009}
Heald G.,  Braun R.,   Edmonds R.,  2009, \mn@doi [\aap]
  {10.1051/0004-6361/200912240}, 503, 409

\bibitem[\protect\citeauthoryear{Heiles \& Haverkorn}{Heiles \&
  Haverkorn}{2012}]{Heiles2012}
Heiles C.,  Haverkorn M.,  2012, \mn@doi [\ssr] {10.1007/s11214-012-9866-4},
  166, 293

\bibitem[\protect\citeauthoryear{{Hill}}{{Hill}}{2018}]{Hill2018}
{Hill} A.,  2018, \mn@doi [Galaxies] {10.3390/galaxies6040129}, \href
  {https://ui.adsabs.harvard.edu/\#abs/2018Galax...6..129H} {6, 129}

\bibitem[\protect\citeauthoryear{{Hill}, {Joung}, {Mac Low}, {Benjamin},
  {Haffner}, {Klingenberg}  \& {Waagan}}{{Hill} et~al.}{2012}]{Hill2012}
{Hill} A.~S.,  {Joung} M.~R.,  {Mac Low} M.-M.,  {Benjamin} R.~A.,  {Haffner}
  L.~M.,  {Klingenberg} C.,   {Waagan} K.,  2012, \mn@doi [\apj]
  {10.1088/0004-637X/750/2/104}, \href
  {https://ui.adsabs.harvard.edu/\#abs/2012ApJ...750..104H} {750, 104}

\bibitem[\protect\citeauthoryear{Hill et~al.,}{Hill et~al.}{2017}]{Hill2017}
Hill A.~S.,  et~al., 2017, \mn@doi [MNRAS in press] {10.1093/mnras/stx389}, pp
  1--17

\bibitem[\protect\citeauthoryear{{Hill}, {Mac Low}, {Gatto}  \&
  {Ib{\'a}{\~n}ez-Mej{\'\i}a}}{{Hill} et~al.}{2018}]{Hill2018a}
{Hill} A.~S.,  {Mac Low} M.-M.,  {Gatto} A.,   {Ib{\'a}{\~n}ez-Mej{\'\i}a}
  J.~C.,  2018, \mn@doi [\apj] {10.3847/1538-4357/aacce2}, \href
  {https://ui.adsabs.harvard.edu/\#abs/2018ApJ...862...55H} {862, 55}

\bibitem[\protect\citeauthoryear{Iacobelli et~al.,}{Iacobelli
  et~al.}{2014}]{Iacobelli2014}
Iacobelli M.,  et~al., 2014, \mn@doi [\aap] {10.1051/0004-6361/201322982}, 566,
  5

\bibitem[\protect\citeauthoryear{Kalberla, Kerp, Haud  \& Haverkorn}{Kalberla
  et~al.}{2017}]{Kalberla2017}
Kalberla P. M.~W.,  Kerp J.,  Haud U.,   Haverkorn M.,  2017

\bibitem[\protect\citeauthoryear{Lallement, Vergely, Valette, Puspitarini, Eyer
   \& Casagrande}{Lallement et~al.}{2014}]{Lallement2014}
Lallement R.,  Vergely J.-L.,  Valette B.,  Puspitarini L.,  Eyer L.,
  Casagrande L.,  2014, \mn@doi [\aap] {10.1051/0004-6361/201322032}, 561, A91

\bibitem[\protect\citeauthoryear{Lallement et~al.,}{Lallement
  et~al.}{2018}]{Lallement2018}
Lallement R.,  et~al., 2018, \mn@doi [\aap] {10.1051/0004-6361/201832832}, 616,
  A132

\bibitem[\protect\citeauthoryear{{Liszt}}{{Liszt}}{2014}]{Liszt2014}
{Liszt} H.,  2014, \mn@doi [\apj] {10.1088/0004-637X/780/1/10}, \href
  {https://ui.adsabs.harvard.edu/\#abs/2014ApJ...780...10L} {780, 10}

\bibitem[\protect\citeauthoryear{Manchester, Hobbs, Teoh  \& Hobbs}{Manchester
  et~al.}{2005}]{Manchester2005}
Manchester R.~N.,  Hobbs G.~B.,  Teoh A.,   Hobbs M.,  2005, \mn@doi [\aj]
  {10.1086/428488}, 129, 1993

\bibitem[\protect\citeauthoryear{McClureGriffiths, Dickey, Gaensler, Green  \&
  Haverkorn}{McClureGriffiths et~al.}{2006}]{McClureGriffiths2006}
McClureGriffiths N.~M.,  Dickey J.~M.,  Gaensler B.~M.,  Green A.~J.,
  Haverkorn M.,  2006, \mn@doi [\apj] {10.1086/508706}, 652, 1339

\bibitem[\protect\citeauthoryear{{Mezger} \& {Henderson}}{{Mezger} \&
  {Henderson}}{1967}]{Mezger1967}
{Mezger} P.~G.,  {Henderson} A.~P.,  1967, \mn@doi [\apj] {10.1086/149030},
  \href {https://ui.adsabs.harvard.edu/\#abs/1967ApJ...147..471M} {147, 471}

\bibitem[\protect\citeauthoryear{{Nord}, {Henning}, {Rand}, {Lazio}  \&
  {Kassim}}{{Nord} et~al.}{2006}]{Nord2006}
{Nord} M.~E.,  {Henning} P.~A.,  {Rand} R.~J.,  {Lazio} T. J.~W.,   {Kassim}
  N.~E.,  2006, \mn@doi [\aj] {10.1086/504407}, \href
  {https://ui.adsabs.harvard.edu/\#abs/2006AJ....132..242N} {132, 242}

\bibitem[\protect\citeauthoryear{{Offner}, {Clark}, {Hennebelle}, {Bastian},
  {Bate}, {Hopkins}, {Moraux}  \& {Whitworth}}{{Offner}
  et~al.}{2014}]{Offner2014}
{Offner} S.~S.~R.,  {Clark} P.~C.,  {Hennebelle} P.,  {Bastian} N.,  {Bate}
  M.~R.,  {Hopkins} P.~F.,  {Moraux} E.,   {Whitworth} A.~P.,  2014, in
  {Beuther} H.,  {Klessen} R.~S.,  {Dullemond} C.~P.,   {Henning} T.,  eds,
  Protostars and Planets VI. p.~53 (\mn@eprint {arXiv} {1312.5326}),
  \mn@doi{10.2458/azu_uapress_9780816531240-ch003}

\bibitem[\protect\citeauthoryear{Oppermann et~al.,}{Oppermann
  et~al.}{2012}]{Oppermann2012}
Oppermann N.,  et~al., 2012, \mn@doi [\aap] {10.1051/0004-6361/201118526}, 542,
  A93

\bibitem[\protect\citeauthoryear{{Oppermann} et~al.,}{{Oppermann}
  et~al.}{2015}]{Oppermann2015}
{Oppermann} N.,  et~al., 2015, \mn@doi [\aap] {10.1051/0004-6361/201423995},
  \href {https://ui.adsabs.harvard.edu/abs/2015A&A...575A.118O} {575, A118}

\bibitem[\protect\citeauthoryear{Padoan \& Nordlund}{Padoan \&
  Nordlund}{2011}]{Padoan2011}
Padoan P.,  Nordlund {\AA}.,  2011, \mn@doi [\apj]
  {10.1088/0004-637X/730/1/40}, 730, 40

\bibitem[\protect\citeauthoryear{Price-Whelan et~al.,}{Price-Whelan
  et~al.}{2018}]{Price-Whelan2018}
Price-Whelan A.~M.,  et~al., 2018, \mn@doi [\aj] {10.3847/1538-3881/aabc4f},
  156, 123

\bibitem[\protect\citeauthoryear{Robitaille \& Bressert}{Robitaille \&
  Bressert}{2012}]{Robitaille2012}
Robitaille T.,  Bressert E.,  2012, Astrophysics Source Code Library, p.
  ascl:1208.017

\bibitem[\protect\citeauthoryear{Robitaille et~al.,}{Robitaille
  et~al.}{2013}]{Robitaille2013}
Robitaille T.~P.,  et~al., 2013, \mn@doi [\aap] {10.1051/0004-6361/201322068},
  558, A33

\bibitem[\protect\citeauthoryear{Robitaille et~al.,}{Robitaille
  et~al.}{2017}]{Robitaille2017}
Robitaille J.-F.,  et~al., 2017, \mn@doi [\mnras] {10.1093/mnras/stx642}, 468,
  2957

\bibitem[\protect\citeauthoryear{Robitaille, Scaife, Carretti, Haverkorn,
  Crocker, Kesteven, Poppi  \& Staveley-Smith}{Robitaille
  et~al.}{2018}]{Robitaille2018}
Robitaille J.~F.,  Scaife A. M.~M.,  Carretti E.,  Haverkorn M.,  Crocker
  R.~M.,  Kesteven M.~J.,  Poppi S.,   Staveley-Smith L.,  2018

\bibitem[\protect\citeauthoryear{Roger, Costain, Landecker  \& Swerdlyk}{Roger
  et~al.}{1999}]{Roger1999}
Roger R.~S.,  Costain C.~H.,  Landecker T.~L.,   Swerdlyk C.~M.,  1999, \mn@doi
  [\aaps] {10.1051/aas:1999239}, 137, 7

\bibitem[\protect\citeauthoryear{Rybicki \& Lightman}{Rybicki \&
  Lightman}{1986}]{Rybicki1986}
Rybicki G.~B.,  Lightman A.~P.,  1986, Radiative Processes in Astrophysics,
  \mn@doi{https://doi.org/10.1002/9783527618170.
}

\bibitem[\protect\citeauthoryear{{Schnitzeler}}{{Schnitzeler}}{2010}]{Schnitzeler2010}
{Schnitzeler} D.~H.~F.~M.,  2010, \mn@doi [\mnras]
  {10.1111/j.1745-3933.2010.00957.x}, \href
  {https://ui.adsabs.harvard.edu/\#abs/2010MNRAS.409L..99S} {409, L99}

\bibitem[\protect\citeauthoryear{Shelton}{Shelton}{2009}]{Shelton2009}
Shelton R.~L.,  2009, \mn@doi [\ssr] {10.1007/s11214-008-9358-8}, 143, 231

\bibitem[\protect\citeauthoryear{Sokoloff, Bykov, Shukurov, Berkhuijsen, Beck
  \& Poezd}{Sokoloff et~al.}{1998}]{Sokoloff1998}
Sokoloff D.~D.,  Bykov a.~a.,  Shukurov A.,  Berkhuijsen E.~M.,  Beck R.,
  Poezd a.~D.,  1998, \mn@doi [\mnras] {10.1046/j.1365-8711.1999.02161.x}, 299,
  189

\bibitem[\protect\citeauthoryear{Su et~al.,}{Su et~al.}{2018}]{Su2018}
Su H.,  et~al., 2018, \mn@doi [\mnras] {10.1093/mnras/sty1732}, 479, 4041

\bibitem[\protect\citeauthoryear{Sun, Reich, Waelkens  \& Ensslin}{Sun
  et~al.}{2007}]{Sun2007b}
Sun X.~H.,  Reich W.,  Waelkens A.,   Ensslin T.,  2007, \mn@doi [\aap]
  {10.1051/0004-6361:20078671}, 477, 573

\bibitem[\protect\citeauthoryear{Sun et~al.,}{Sun et~al.}{2015}]{Sun2015}
Sun X.~H.,  et~al., 2015, \mn@doi [\apj] {10.1088/0004-637X/811/1/40}, 811, 40

\bibitem[\protect\citeauthoryear{Taylor, Stil  \& Sunstrum}{Taylor
  et~al.}{2009}]{Taylor2009}
Taylor A.~R.,  Stil J.~M.,   Sunstrum C.,  2009, \mn@doi [\apj]
  {10.1088/0004-637X/702/2/1230}, 702, 1230

\bibitem[\protect\citeauthoryear{Tribble}{Tribble}{1991}]{Tribble1991}
Tribble P.~C.,  1991, \mn@doi [\mnras] {10.1093/mnras/250.4.726}, 250, 726

\bibitem[\protect\citeauthoryear{{Tritsis}, {Federrath}  \&
  {Pavlidou}}{{Tritsis} et~al.}{2019}]{Tritsis2018}
{Tritsis} A.,  {Federrath} C.,   {Pavlidou} V.,  2019, \mn@doi [\apj]
  {10.3847/1538-4357/ab037d}, \href
  {https://ui.adsabs.harvard.edu/abs/2019ApJ...873...38T} {873, 38}

\bibitem[\protect\citeauthoryear{Uyaniker, Landecker, Gray  \& Kothes}{Uyaniker
  et~al.}{2003}]{Uyaniker2003}
Uyaniker B.,  Landecker T.~L.,  Gray A.~D.,   Kothes R.,  2003, \mn@doi [\apj]
  {10.1086/346234}, 585, 785

\bibitem[\protect\citeauthoryear{{Van Eck} et~al.,}{{Van Eck}
  et~al.}{2017}]{VanEck2017}
{Van Eck} C.~L.,  et~al., 2017, \mn@doi [\aap] {10.1051/0004-6361/201629707},
  597, A98

\bibitem[\protect\citeauthoryear{Vergely, Valette, Lallement  \&
  Raimond}{Vergely et~al.}{2010}]{Vergely2010a}
Vergely J.-L.,  Valette B.,  Lallement R.,   Raimond S.,  2010, \mn@doi [\aap]
  {10.1051/0004-6361/200913962}, 518, A31

\bibitem[\protect\citeauthoryear{{Wardle} \& {Sramek}}{{Wardle} \&
  {Sramek}}{1974}]{Wardle1974}
{Wardle} J.~F.~C.,  {Sramek} R.~A.,  1974, \mn@doi [\apj] {10.1086/152816},
  \href {https://ui.adsabs.harvard.edu/abs/1974ApJ...189..399W} {189, 399}

\bibitem[\protect\citeauthoryear{{Wolleben} et~al.,}{{Wolleben}
  et~al.}{2009}]{Wolleben2008}
{Wolleben} M.,  et~al., 2009, in {Strassmeier} K.~G.,  {Kosovichev} A.~G.,
  {Beckman} J.~E.,  eds,  IAU Symposium Vol. 259, Cosmic Magnetic Fields: From
  Planets, to Stars and Galaxies. pp 89--90 (\mn@eprint {arXiv} {0812.2450}),
  \mn@doi{10.1017/S1743921309030117}

\bibitem[\protect\citeauthoryear{Wolleben, Landecker, Hovey, Messing, Davison,
  House, Somaratne  \& Tashev}{Wolleben et~al.}{2010a}]{Wolleben2010a}
Wolleben M.,  Landecker T.~L.,  Hovey G.~J.,  Messing R.,  Davison O.~S.,
  House N.~L.,  Somaratne K. H. M.~S.,   Tashev I.,  2010a, \mn@doi [\aj]
  {10.1088/0004-6256/139/4/1681}, 139, 1681

\bibitem[\protect\citeauthoryear{Wolleben et~al.,}{Wolleben
  et~al.}{2010b}]{Wolleben2010}
Wolleben M.,  et~al., 2010b, \mn@doi [\apj] {10.1088/2041-8205/724/1/L48}, 724,
  L48

\bibitem[\protect\citeauthoryear{{Wood}, {Haffner}, {Reynolds}, {Mathis}  \&
  {Madsen}}{{Wood} et~al.}{2005}]{Wood2005}
{Wood} K.,  {Haffner} L.~M.,  {Reynolds} R.~J.,  {Mathis} J.~S.,   {Madsen} G.,
   2005, \mn@doi [\apj] {10.1086/432831}, \href
  {https://ui.adsabs.harvard.edu/\#abs/2005ApJ...633..295W} {633, 295}

\bibitem[\protect\citeauthoryear{Zheng et~al.,}{Zheng et~al.}{2017}]{Zheng2017}
Zheng H.,  et~al., 2017, \mn@doi [\mnras] {10.1093/mnras/stw2525}, 464, 3486

\bibitem[\protect\citeauthoryear{van Leeuwen}{van
  Leeuwen}{2007}]{VanLeeuwen2007}
van Leeuwen F.,  2007, \mn@doi [\aap] {10.1051/0004-6361:20078357}, 474, 653

\makeatother
\end{thebibliography}




\appendix

\section{Faraday Thick Spectra}\label{sec:appthick}
We model Stokes $Q$ and $U$, and PI as a function of $\lambda^2$ in the GMIMS-LBS band using Equation~\ref{eqn:slab}. Models 1.X, 2.X, 3.X refer to Faraday thick cases (1), (2), and (3) as described in Section~\ref{sec:modelcube}. The `X' value for each model refers to which $\chi_0$ value is used for each slab. This is because there is a choice as to which $\chi_0$ value to use from the two peaks which become the edges of the slab. We set the height of each Burn slab to be 0.25\,K/\radms\ to give a resolved height of about 0.18\,K/RMSF. In all cases, the fit to the original data is poor.

\begin{figure*}
	\centering
	\includegraphics[width=\linewidth]{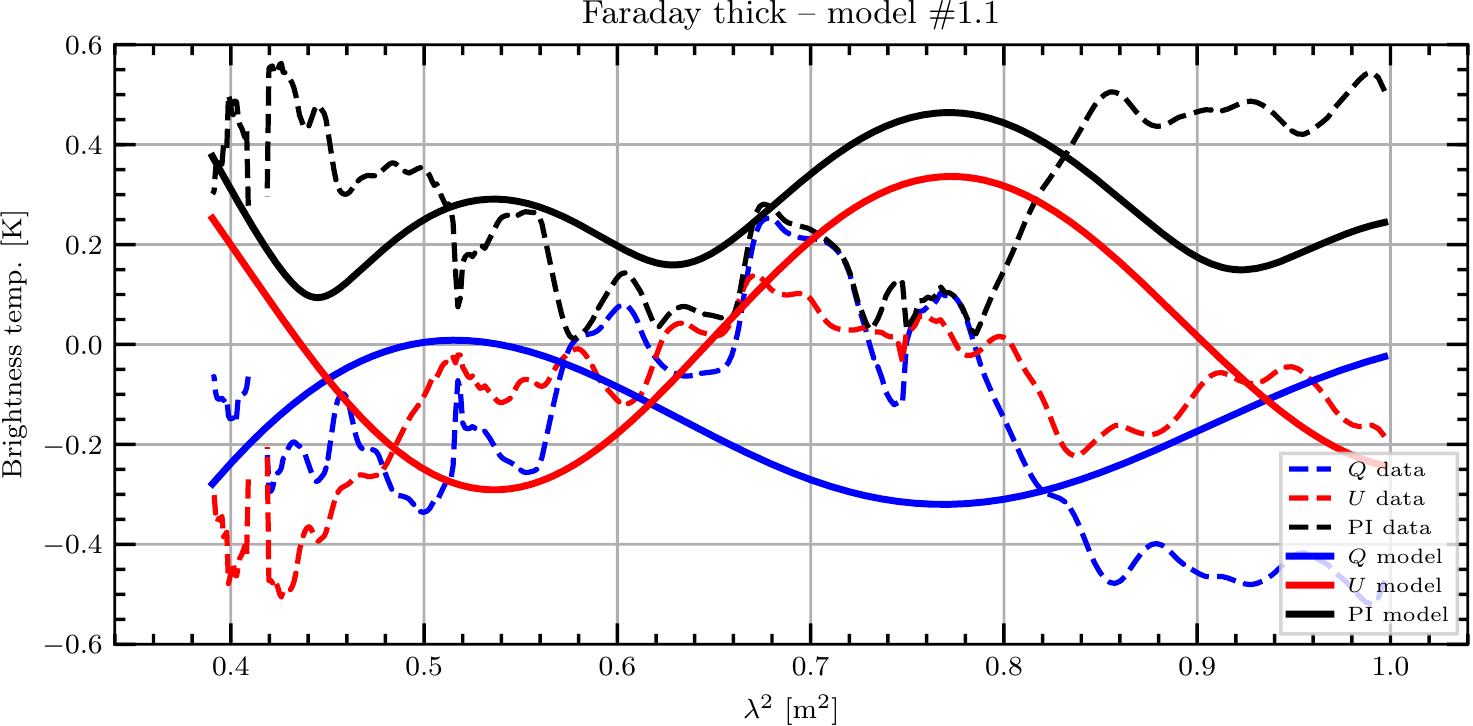}
	\caption{Faraday thick model spectra towards Sh2-27: A Burn slab spanning peaks 1 and 2, taking $\chi_0$ from peak 1, and a Faraday thin component at peak 3. Dashed lines: Average Stokes $Q$, $U$, and PI $\lambda^2$ spectra towards Sh-27 from GMIMS-LBS. Solid lines: Faraday thick model derived from the average Faraday spectrum.}
	\label{fig:thick1_1}
\end{figure*}
\begin{figure*}
	\centering
	\includegraphics[width=\linewidth]{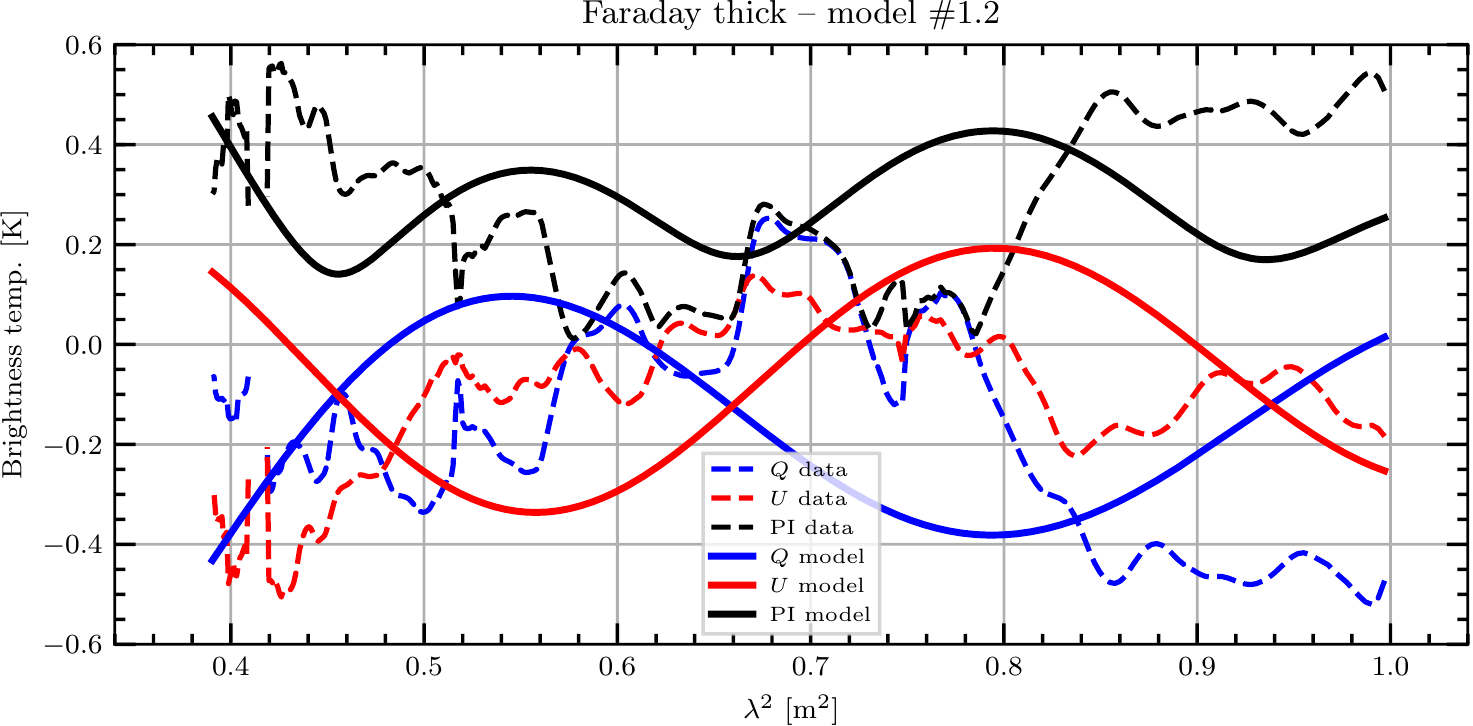}
	\caption{Faraday thick model spectra towards Sh2-27: A Burn slab spanning peaks 1 and 2, taking $\chi_0$ from peak 2, and a Faraday thin component at peak 3. Dashed lines: Average Stokes $Q$, $U$, and PI $\lambda^2$ spectra towards Sh-27 from GMIMS-LBS. Solid lines: Faraday thick model derived from the average Faraday spectrum.}
	\label{fig:thick1_2}
\end{figure*}
\begin{figure*}
	\centering
	\includegraphics[width=\linewidth]{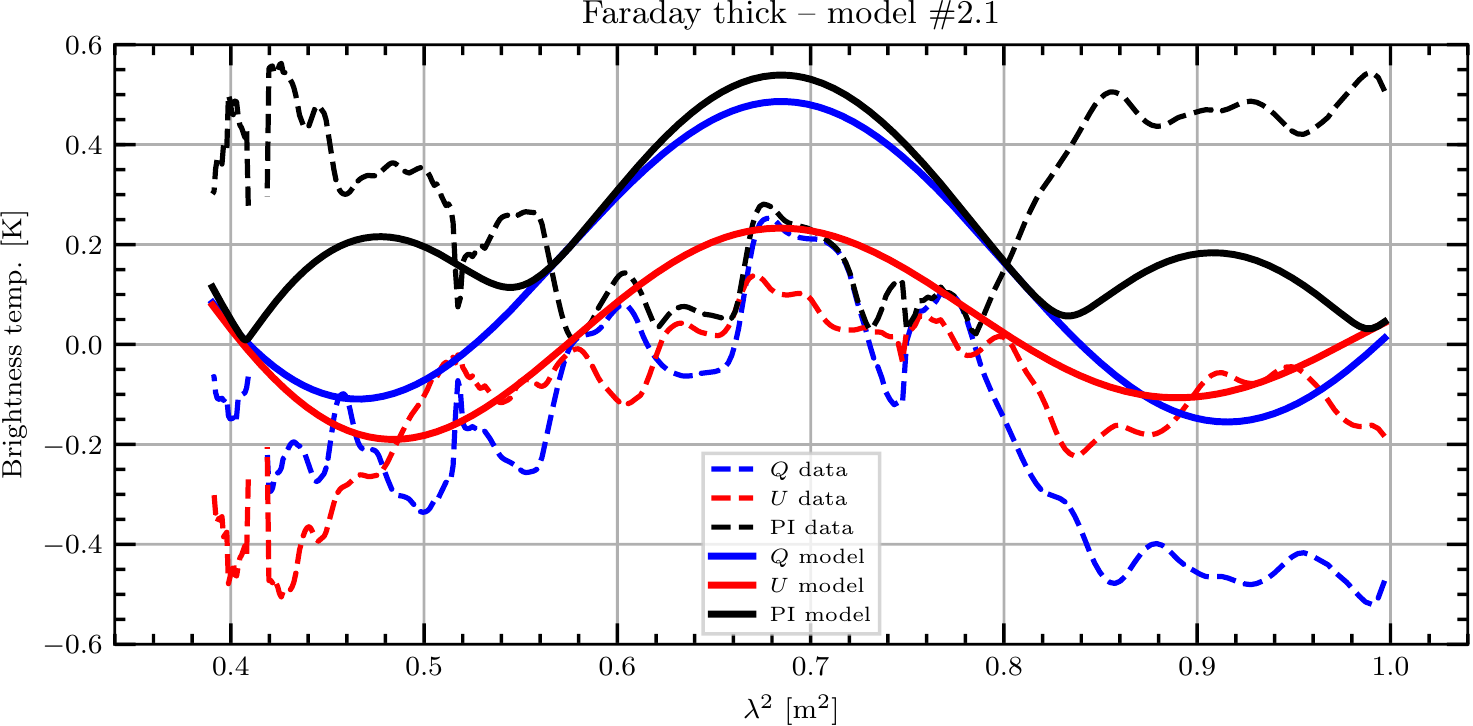}
	\caption{Faraday thick model spectra towards Sh2-27: A Burn slab spanning peaks 2 and 3, taking $\chi_0$ from peak 3, and a Faraday thin component at peak 1. Dashed lines: Average Stokes $Q$, $U$, and PI $\lambda^2$ spectra towards Sh-27 from GMIMS-LBS. Solid lines: Faraday thick model derived from the average Faraday spectrum.}
	\label{fig:thick2_1}
\end{figure*}
\begin{figure*}
	\centering
	\includegraphics[width=\linewidth]{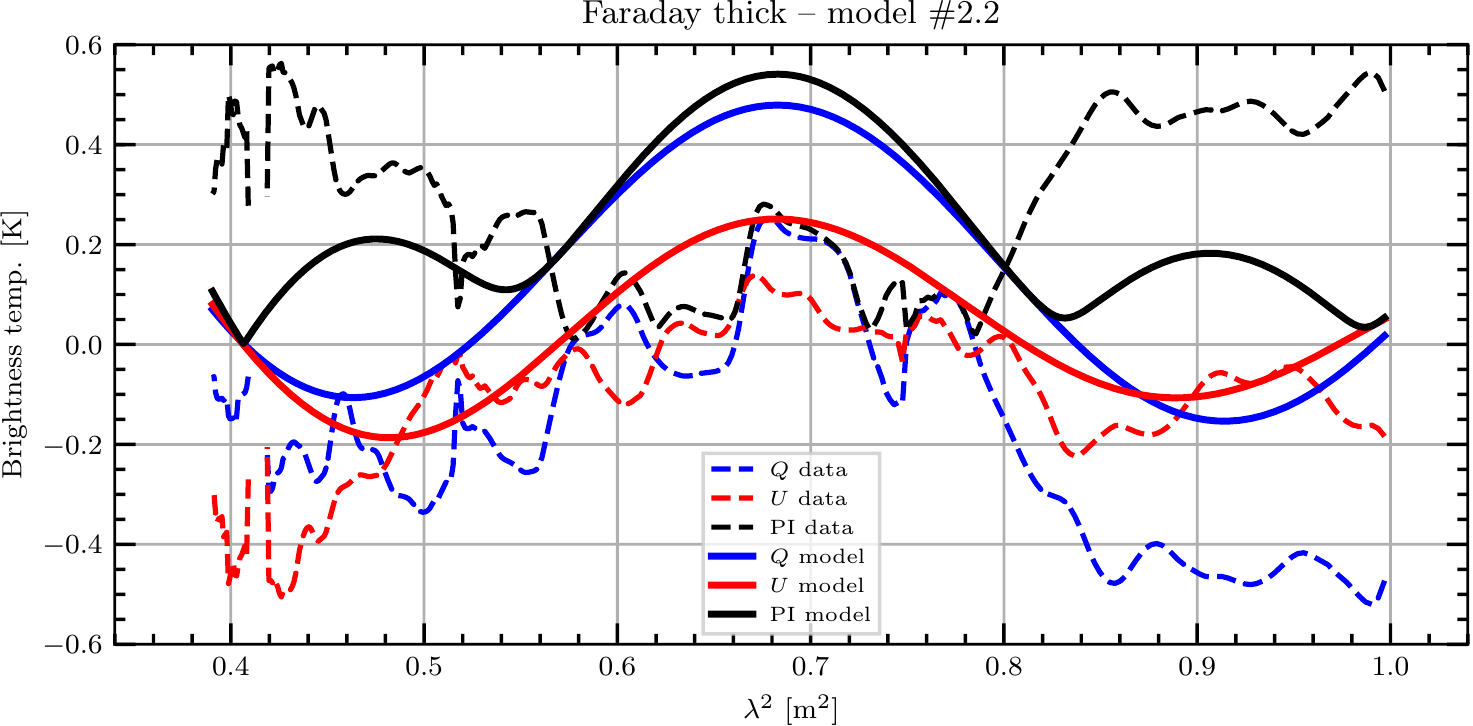}
	\caption{Faraday thick model spectra towards Sh2-27: A Burn slab spanning peaks 2 and 3, taking $\chi_0$ from peak 2, and a Faraday thin component at peak 1. Dashed lines: Average Stokes $Q$, $U$, and PI $\lambda^2$ spectra towards Sh-27 from GMIMS-LBS. Solid lines: Faraday thick model derived from the average Faraday spectrum.}
	\label{fig:thick2_2}
\end{figure*}
\begin{figure*}
	\centering
	\includegraphics[width=\linewidth]{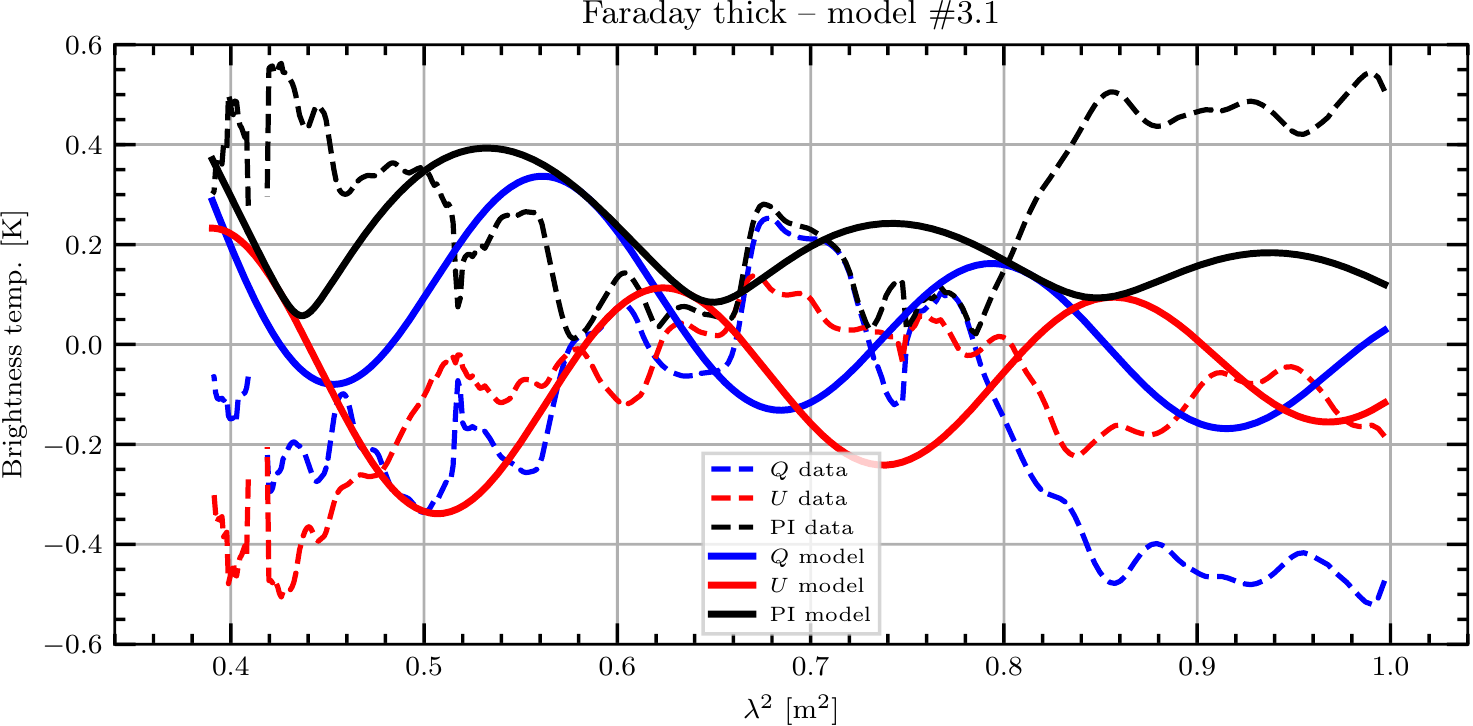}
	\caption{Faraday thick model spectra towards Sh2-27: A Burn slab spanning peaks 1 and 3, taking $\chi_0$ from peak 3, and a Faraday thin component at peak 2. Dashed lines: Average Stokes $Q$, $U$, and PI $\lambda^2$ spectra towards Sh-27 from GMIMS-LBS. Solid lines: Faraday thick model derived from the average Faraday spectrum.}
	\label{fig:thick3_1}
\end{figure*}
\begin{figure*}
	\centering
	\includegraphics[width=\linewidth]{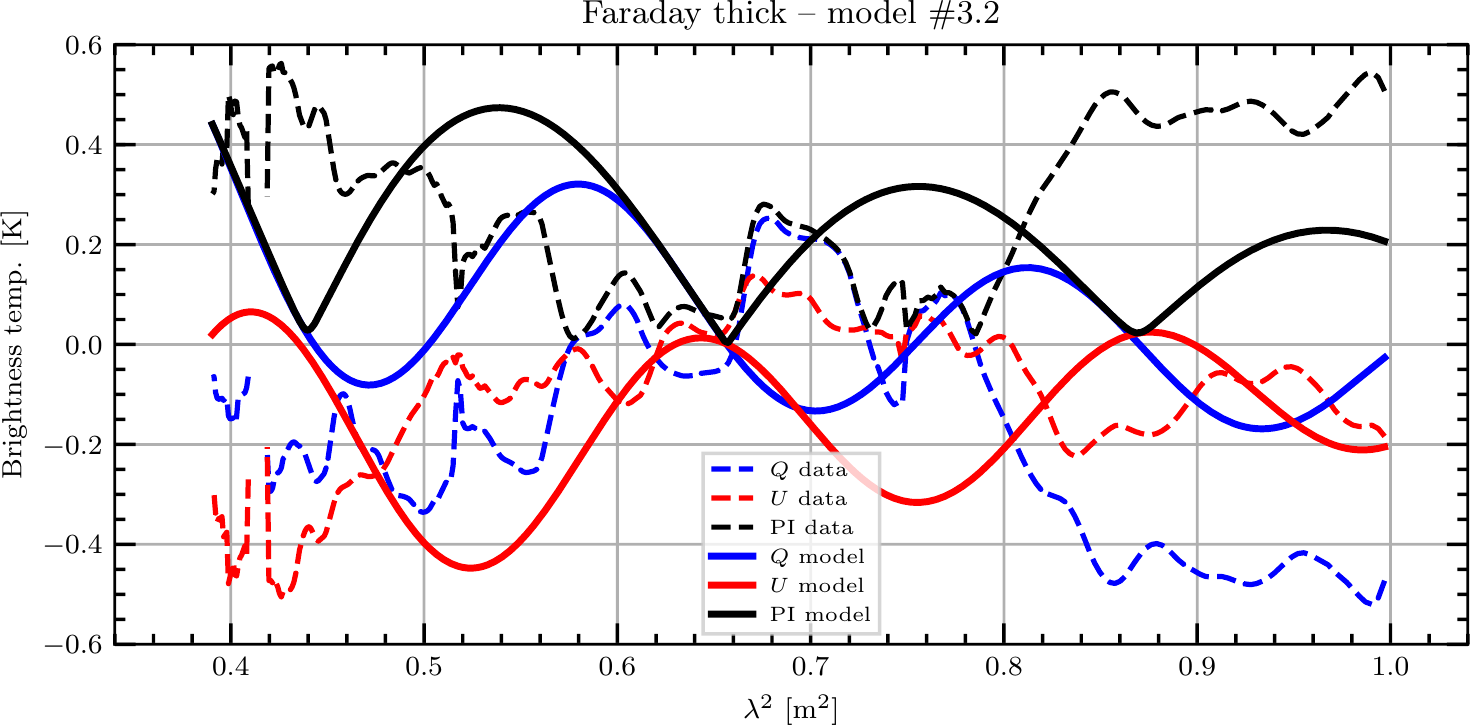}
	\caption{Faraday thick model spectra towards Sh2-27: A Burn slab spanning peaks 1 and 3, taking $\chi_0$ from peak 1, and a Faraday thin component at peak 2. Dashed lines: Average Stokes $Q$, $U$, and PI $\lambda^2$ spectra towards Sh-27 from GMIMS-LBS. Solid lines: Faraday thick model derived from the average Faraday spectrum.}
	\label{fig:thick3_2}
\end{figure*}


\bsp	
\label{lastpage}
\end{document}